\documentclass[11pt]{amsart}
\usepackage{amsaddr}
\usepackage{graphicx} % Required for inserting images
\usepackage{amsmath}
\usepackage{amssymb}
\usepackage{mathpazo}
\usepackage{hyperref}
\usepackage{etoolbox}
\usepackage{setspace}

\usepackage[most]{tcolorbox}

\numberwithin{equation}{section}

\newcommand{\bea}{\begin{eqnarray}\displaystyle}
\newcommand{\eea}{\end{eqnarray}}

\newcommand{\del}{\partial}

\newcommand{\ov}{\overline}

\newcommand{\be}{\begin{equation}}
\newcommand{\ee}{\end{equation}}

\usepackage{amsthm}

\newtheorem{conjecture}{Conjecture}
\newtheorem{theorem}{Theorem}[section] % Theorems numbered within sections
\theoremstyle{definition} % This style will italicize the body

\newtheorem{remark}[theorem]{Remark}

\title{Poisson Vertex Algebra of Seiberg-Witten Theory}
\author{Ahsan Z. Khan }
\address{Harvard University,\\
         Center for Mathematical Sciences and Applications,\\
         Cambridge, MA 02138}
\email{ahsankhan@g.harvard.edu}
\date{\today}
\begin{document}

\maketitle

\begin{abstract} The space of local operators in the $Q$-cohomology of the holomorphic-topological supercharge in a four-dimensional $\mathcal{N}=2$ theory carries the structure of a Poisson vertex algebra. This note studies the  Poisson vertex algebra associated to the pure $\mathcal{N}=2$ gauge theory with gauge group $SU(2)$. We propose an explicit Poisson vertex algebra $A$, claimed to be isomorphic to the algebra of holomorphic-topological observables to all orders in perturbation theory. We compute the Hilbert-Poincar\'e series of $A$ and show that it refines the Schur index of the pure $SU(2)$ theory. We show that $A$ admits a further differential $Q_{\text{inst}}$ which we hypothesize captures non-perturbative corrections, and compute the cohomology of this differential. We thus present an explicit candidate for the space of non-perturbative holomorphic-topological observables of Seiberg-Witten theory.  
\end{abstract}

\section{Introduction}  
The study of local observables in supersymmetric field theories has repeatedly led to highly rich mathematical structures: quantum cohomology \cite{Vafa:1991uz}, $A_{\infty}$ categories of boundary conditions \cite{Gaiotto:2015aoa}, vertex operator algebras \cite{Witten:2005px, Beem:2013sza} and quantum groups \cite{Costello:2013zra} to name a few. A particularly striking instance of this phenomenon occurs in four-dimensional theories with $\mathcal{N}=2$ supersymmetry. In the case of superconformal theories, it was shown in \cite{Beem:2013sza} that a certain $Q$-cohomology of local operators forms a vertex operator algebra. This remarkable observation led to an explosion of activity, resulting in many subsequent developments\footnote{For an account of some of the more recent developments see \cite{Dedushenko:2023cvd} and references therein.}. 

The vertex operator algebra associated to a four-dimensional $\mathcal{N}=2$ superconformal field theory admits a natural reinterpretation in terms of the holomorphic-topological twist\footnote{The holomorphic-topological twist of four-dimensional $\mathcal{N}=2$ theories is sometimes also called the Kapustin twist \cite{Kapustin:2006hi}.} \cite{Jeong:2019pzg, Oh:2019bgz, Butson:2020mmu}. From this perspective, the vertex algebra arises from quantizing a more primitive structure: the algebra of local observables in the cohomology of the holomorphic-topological supercharge. Importantly, the holomorphic topological twist makes sense for any four-dimensional $\mathcal{N}=2$ theory, not only in the superconformal case, and thus provides a framework for studying protected operator algebras more generally.

Concretely, consider a four-dimensional $\mathcal{N}=2$ theory on $\mathbb{R}^4 = \mathbb{R}^2 \times \mathbb{C}$ with coordinates $(w,\bar w,z,\bar z)$ and denote the holomorphic-topological supercharge as $Q_{\mathrm{HT}}$. This supercharge renders the translations $\partial_w$, $\partial_{\bar w}$, and $\partial_{\bar z}$ as null-homotopic, while leaving the action of the holomorphic translation $\partial_z$ as nontrivial. In such a situation one expects the space of local operators in $Q_{\mathrm{HT}}$-cohomology to carry the structure of a Poisson vertex algebra\footnote{The cochain level lift of this structure is sometimes called an $(E_2,\, \text{chiral})$ algebra. See for instance \cite{Dimofte:2025oqf}.} \cite{Oh:2019mcg, Butson:2020coe}. The vertex operator algebra of \cite{Beem:2013sza} is then obtained from this classical Poisson vertex algebra by quantizing via an $\Omega$-deformation \cite{Jeong:2019pzg, Oh:2019bgz}, the latter of which is available only in the superconformal case.

A Poisson vertex algebra $V$ consists of a derivation $\partial$, a graded-commutative product, and a $\lambda$-bracket $\{\cdot_\lambda \cdot\}$ satisfying compatibility axioms. In a holomorphic-topological field theory on $\mathbb{R}^2 \times \mathbb{C}$, the derivation $\partial$ corresponds to the holomorphic derivative of operators, the commutative product arises from the nonsingular operator product expansion (nonsingular due to the presence of topological directions), and the $\lambda$-bracket encodes a secondary operation \cite{Beem:2018fng}, taking the schematic form 
\begin{align}
    \{\mathcal{O}_1 \,_{\lambda} \,\mathcal{O}_2 \} = \oint \text{d}z\, e^{\lambda z} \int_{\mathbb{R}^2_{(x,y)}} (\mathcal{O}_1)^{(2)}(x, y, z) \,\mathcal{O}_2(0).
\end{align}

While general arguments establish the existence of this Poisson vertex algebra structure, its explicit form is poorly understood for generic four-dimensional $\mathcal{N}=2$ theories. As we will explain, for Lagrangian $\mathcal{N}=2$ theories, one can write down a BRST cochain complex whose cohomology computes the algebra of observables, but determining this cohomology is quite nontrivial\footnote{For previous work on holomorphic-topological observables in Lagrangian $\mathcal{N}=2$ theories, specifically in connection to the affine Grassmannian see \cite{Niu:2021jet}. The latter reference builds on a remarkable mathematical proposal for the category of line defects in the holomorphic-topological twist of Lagrangian $\mathcal{N}=2$ theories \cite{cw}. }.

The purpose of this note is to address this problem in the simplest non-trivial, and perhaps most extensively studied example: pure four-dimensional $\mathcal{N}=2$ gauge theory with gauge group $SU(2)$, originally studied by Seiberg and Witten \cite{Seiberg:1994rs}. Our main contributions can be summarized as follows:
\begin{itemize}
    \item We define a Poisson vertex algebra $A$ as being generated by an even Virasoro element $X$ with central charge $c=0$, and an odd field $Y$ carrying conformal weight $3$ under $X$, and quotienting by the smallest Poisson vertex ideal containing the element $X^2$. We compute the Hilbert-Poincar\'e series of $A$, which organizes the tri-graded dimensions of $A$. The resulting series \begin{align}
        P_A(t,q,y) = \sum_{n = 0}^{\infty} q^{n(n+1)} y^n t^{2n} \frac{(-t y^2 q ; q)_n}{(q;q)_n},
    \end{align} gives a bigraded refinement of the Schur index of the pure $\mathcal{N}=2$, $SU(2)$ gauge theory. \\ 
    \item We formulate the algebra of perturbative holomorphic-topological observables $\text{Obs}_{\text{HT}}(\mathfrak{g})$ of an $\mathcal{N}=2$ $\mathfrak{g}$-valued vector multiplet as the cohomology of a differential-graded Poisson vertex algebra of \textit{basic} elements of a $\mathfrak{g}$-valued $bc$-ghost system, with respect to the natural BRST differential. As a graded vector space, this corresponds to the Lie algebra cohomology\footnote{An equivalent way to write this is as the Lie algebra cohomology of the (even) Lie algebra $\mathfrak{g}[z, \varepsilon] = \mathfrak{g}[[z]] \ltimes \varepsilon\, \mathfrak{g}[[z]] $  where $\varepsilon$ is an \textit{even} square zero parameter $\varepsilon^2 = 0$, relative to the Lie subalgebra $\mathfrak{g} \subset \mathfrak{g}[z, \varepsilon]$: \begin{align}
        \text{Obs}_{\text{HT}}(\mathfrak{g}) = H^*(\mathfrak{g}[z, \varepsilon], \mathfrak{g}).
    \end{align} In the above we've written $\varepsilon = \text{d}z$.} of the Lie algebra $\mathfrak{g}[[z]]$, relative to the Lie subalgebra $\mathfrak{g} \subset \mathfrak{g}[[z]]$, with values in the $\mathfrak{g}[[z]]$-module given by $\Lambda((\mathfrak{g}[[z]] \,\text{d}z)^{\vee}):$ 
    \begin{align}
        \text{Obs}_{\text{HT}}(\mathfrak{g}) = H^*(\mathfrak{g}[[z]], \mathfrak{g} \,; \, \Lambda\big((\mathfrak{g}[[z]] \,\text{d}z)^{\vee}) \big). 
    \end{align} $$ $$
    \item For $\mathfrak{g}=\mathfrak{sl}_2$, we construct a map of Poisson vertex algebras \begin{align} \varphi: A \to \text{Obs}_{\text{HT}}(\mathfrak{\mathfrak{sl}_2}) ,\end{align} and conjecture that the map $\varphi$ is an isomorphism. The claim that $\varphi$ is an isomorphism is supported by extensive Euler characteristic checks, along with a direct computational linear algebra calculation of the relative Lie algebra cohomology for spins $S \leq 20,$ agreeing with the exact Hilbert-Poincar\'e series $P_A(t,q,y)$ to the given order. \\
    \item Finally, we define a differential $Q_{\text{inst}}$ on $A$ itself that we expect captures the effect of non-perturbative corrections to the BRST differential, and compute the cohomology of $Q_{\text{inst}}$ exactly. We find that there is a single surviving operator $\alpha_{2n}$ at each non-negative even ghost number taking the form 
    \begin{align}
        \alpha_{2n} = [X \,\del^2 X \dots \del^{2n-2}X], \,\,\,\,\,\,\, n \geq 0,
    \end{align} carrying spin $s_n = n(n+1),$ resulting in the Hilbert-Poincar\'e series 
    \begin{align}
      P_{H^*(A)}(t,q) = \sum_{n =0}^{\infty} t^{2n} q^{n(n+1)}.    
    \end{align} 

\end{itemize}

The organization of this note is as follows. In Section \ref{PVA} we define the Poisson vertex algebra $A$ and compute its Hilbert-Poincar\'e series. In Section \ref{sw} we formulate the holomorphic-topological twist of an $\mathcal{N}=2$ vector multiplet, and the algebra of holomorphic topological observables $\text{Obs}_{\text{HT}}(\mathfrak{g})$ as a complex of basic invariants in the $bc$-ghost system. For the case of $\mathfrak{g}=\mathfrak{sl}_2$ we then construct a map of Poisson vertex algebras from $A$ to the algebra of holomorphic-topological observables.  In Section \ref{np} we introduce the differential $Q_{\mathrm{inst}}$ on $A$, a differential hypothesized to capture non-perturbative effects, and compute its cohomology. We conclude in Section \ref{future} with a discussion of future directions. 

\textbf{Remark on AI Usage:} Interaction with a frontier large language model (\texttt{GPT 5.2-Pro}) proved to be quite important during the exploratory stage of this project. A brief methodological account of this usage is included in Appendix \ref{gpt}. We emphasize that while AI was used as an exploratory tool, the paper is entirely human authored. In particular, the author assumes responsibility of any errors.

\section*{Acknowledgements} I thank Davide Gaiotto for collaboration at the initial stages of the project and for useful feedback. I'm also grateful to Kasia Budzik for doing some finite spin calculations at an early stage of the project. A special thanks goes to Sunghyuk Park for pointing out a key $q$-series formula. Finally, I thank Greg Moore for stimulating discussions, and Anindya Banerjee, Kevin Costello, Michael Douglas, and Constantin Teleman for helpful correspondence. 

This work is supported by the Center for Mathematical Sciences and Applications at Harvard University.

\section{The Poisson Vertex Algebra $A$} \label{PVA} \subsection{Background} 

Recall that a super Poisson vertex algebra $V$ is a $\mathbb{Z}_2$-graded vector space
\begin{align}
V = V^0\oplus V^1,
\end{align}
equipped with a supercommutative product $\cdot : V\otimes V \to V,$
and an even derivation $\partial: V\to V$ of this product, so that
\be \partial(ab) = (\partial a)b + a(\partial b). \ee In addition, $V$ is equipped with an even $\lambda$-bracket,
\begin{align}
\{\cdot \,_{\lambda} \cdot \} : V \otimes V \to \mathbb{C}[\lambda]\otimes V,
\end{align}
written as
\begin{align}
\{a \,_{\lambda} \,b\} = \sum_{n \geq 0} \lambda^n (a_{(n)} b) , 
\qquad 
(a_{(n)} b )\in V,
\end{align}
subject to the axioms of conformal sesquilinearity,
\begin{align}
\{\partial a_\lambda b\} &= -\lambda \{a_\lambda b\}, 
\qquad
\{a_\lambda \partial b\} = (\partial + \lambda)\{a_\lambda b\}.
\end{align}
skew-symmetry,
\begin{equation}
\{a_\lambda b\}
=
-(-1)^{p(a)p(b)} 
\{\, b_{-\lambda-\partial} a \},
\end{equation}
and the $\lambda$-bracket Jacobi identity,
\begin{equation}
\{a_\lambda \{b_\mu c\}\}
=
\{\{a_\lambda b\}_{\lambda+\mu} c\}
+
(-1)^{p(a)p(b)}
\{b_\mu \{a_\lambda c\}\},
\end{equation} where $p(a)$ denotes the parity of a homogenous element $a\in V$.

The commutative product and $\lambda$-bracket are moreover required to be compatible:
\begin{align} 
\label{leib1} \{a_\lambda bc\}
&=
\{a_\lambda b\}c
+
(-1)^{p(a)p(b)} b\{a_\lambda c\}, \\
 \label{leib2} \{ab_\lambda c\}
&=
(-1)^{p(b) p(c)}\{a_{\lambda+\partial} c\}_{\to}b
+
(-1)^{p(a)(p(b)+p(c))}
\{b_{\lambda+\partial} c\}_{\to}a.
\end{align} Henceforth we will drop the ``super" descriptor and simply refer to the above structure as a Poisson vertex algebra. 

A $\mathbb{Z}_2$-graded subspace $I \subset V$ is called a Poisson vertex ideal if $I$ is a differential ideal:
\begin{align}
\partial I \subset I,
\qquad
VI \subset I,
\end{align}
and $I$ is closed under the $\lambda$-bracket:
\begin{align}
\{V\,{_\lambda} \,I\} \subset I[\lambda]
\end{align}
The quotient $V/I$ inherits the structure of a Poisson vertex algebra.

For more thorough reviews of Poisson vertex algebras and their connection to various subjects in mathematical physics, we refer the reader to \cite{kacreview, Khan:2025rah, Oh:2019mcg}. 

In the Poisson vertex algebras arising in four-dimensional $\mathcal{N}=2$ field theories, $V$ carries two additional $\mathbb{Z}$-gradings. A \emph{ghost number} grading, denoted $g$, and a \emph{spin} grading, denoted $s$. We therefore write
\begin{align}
V = \bigoplus_{(s,g)} V^{(s,g)}.
\end{align}
The product preserves both gradings, whereas the derivation $\partial$ preserves ghost number and raises spin by one:
\begin{align}
\partial : V^{(s,g)} \to V^{(s+1,g)}.
\end{align}
The $\lambda$-bracket is \textit{shifted} in the sense that it carries ghost number $-2$, \be \{ \cdot \,_{\lambda} \cdot \}: V^{g_1} \otimes V^{g_2} \to V^{g_1 + g_2 - 2}[\lambda].\ee Moreover, assigning the formal variable $\lambda$ to carry spin $1$, the $\lambda$-bracket carries spin $-1$. All-in-all, these gradings can be summarized by saying that the $n$th coefficient map in the $\lambda$-bracket is a map
\begin{align}
(\cdot_{(n)} \cdot)
:
V^{(s_1,g_1)} \otimes V^{(s_2,g_2)}
\longrightarrow
V^{(s_1 + s_2 - (n+1),\, g_1 + g_2 - 2)}.
\end{align}

Let us now come to the description of the Poisson vertex algebra of interest. 

\subsection{The Poisson Vertex Algebra $A$}

Let $V$ be the Poisson vertex algebra defined as follows. As a differential supercommutative algebra, it is generated by a pair of fields $X, Y$, where $X$ is even and $Y$ is odd, so that as a vector space \be V = \mathbb{C}[X, \del X, \del^2 X, \dots ] \otimes \Lambda[ Y, \del Y, \del^2 Y, \dots].\ee The derivation $\del$ acts in the obvious way, and the product is the free graded commutative product where $X$ is commutative, and $Y$ is anti-commutative: \be \del^k X \,\del^l X = \del^l X \,\del^k X, \,\,\,\,\, \del^k Y \,\del^l X = \del^l X\, \del^k Y, \,\,\,\,\,\, \del^k Y \, \del^l Y = - \del^l Y \,\del^k Y.\ee Since $V$ is a free supercommutative differential algebra, to specify the $\lambda$-bracket on $V$, it suffices to specify it on generators, and check the Jacobi identity on those only. We set 
\begin{align}
    \{X \,_{\lambda} \,X\} &= \del X + 2 \lambda X, \\ \{X \,_{\lambda} \, Y\} &= \del Y + 3 \lambda Y, \\ \{Y\,_{\lambda}\, Y\} &= 0. 
\end{align} The shape of these $\lambda$-brackets is in fact standard. The $\lambda$-bracket of $X$ with itself says that $X$ is a Virasoro element of central charge $c=0$, whereas the $\lambda$-bracket of $X$ with $Y$ says that $Y$ carries conformal weight $3$ under the Virasoro element $X$. 

To specify the additional spin and ghost number gradings, we set both the spin and ghost number of $X$ to be $+2$, whereas the spin and ghost number of $Y$ are both set to be $+3$, so that 
\begin{align}
s(\del^kX) &= k+2, \quad s(\del^kY) = k+3, \\ g(\del^k X) &= 2, \,\,\,\,\,\,\,\,\,\,\quad g(\del^kY) = 3. 
\end{align} With these gradings the $\lambda$-bracket indeed carries ghost number $-2$ and spin $-1$.

Define $I \subset V$ to be the smallest Poisson vertex ideal of $V$ containing the element $X^2.$ The main Poisson vertex algebra of interest in this note is the quotient \be A(\mathfrak{sl}_2) := V/I,\ee equipped with the induced Poisson vertex structure. The Lie algebra $\mathfrak{sl}_2$ arising in the notation will be explained later. Since $\mathfrak{sl}_2$ is the only Lie algebra that will be considered in detail in this note, we henceforth drop it from the notation and simply denote the quotient Poisson vertex algebra as $A$.
 
We now describe $A$ as a differential supercommutative algebra.

Since $I$ is a Poisson vertex ideal, it must be closed under $\del$ and should be an invariant subspace with respect to the adjoint $\lambda$-bracket map. Starting from $X^2$, taking the $\lambda$-bracket with respect to $Y$ gives \be \{Y \,_{\lambda} \, X^2 \} = 4 X \del Y + 6 XY \lambda . \ee This element is in $I[\lambda]$ if and only if \be X Y, X \del Y \in I.\ee Moreover taking $\lambda$-bracket of $XY$ with $Y$ must give an element of $I[\lambda]$. Since $Y^2=0$, this yields \be Y \del Y \in I. \ee 
Let
\begin{align}
J := \langle X^2,\; XY,\; X\partial Y,\; Y\partial Y\rangle_{\partial}
\subset V
\end{align}
denote the differential ideal generated by these four elements. We claim that $J$ is a Poisson vertex ideal. Since $J$ contains $X^2$, this will imply $I\subset J$ by minimality of $I$. On the other hand the computations above show $X^2,XY,X\partial Y,Y\partial Y\in I$, hence $J\subset I$. Therefore $I=J$, and no further generators occur.

Showing that $J$ is a Poisson vertex ideal is a direct computation. One simply has to compute $\{X\,_{\lambda}\, a\}$ and $\{Y\,_{\lambda}\, a \}$ and show that they reside in $J[\lambda]$ for $a$ being each of the four generators of $J$. This can be checked by applying the standard PVA axioms. For instance we have \begin{align}
\{X\,_{\lambda}\, X^2\} &= \del(X^2) +   4X^2 \lambda , \\
\{X \,_{\lambda}\, XY \} &= \del(XY) + 5XY \lambda, \\ 
\{X \,_{\lambda}\, X \del Y \} &= \del(X \del Y) +  6 X \del Y \lambda + 3 XY \lambda^2, \\ 
\{X \,_{\lambda}\, Y \del Y \} &= \del(Y \del Y) + 7 Y \del Y \lambda.
\end{align} Similarly, one can check closure under $\{Y\,_{\lambda} \, \cdot \}.$
Thus
\be
I
=
\langle X^2,\; XY,\; X\partial Y,\; Y\partial Y\rangle_{\partial}.
\ee Thus as a differential supercommutative algebra, \be A = \mathbb{C}[X, \del X, \dots]\otimes \Lambda[Y, \del Y, \dots]/\langle X^2,\; XY,\; X\partial Y,\; Y\partial Y\rangle_{\partial} . \ee

\subsection{Hilbert-Poincar\'e Series of $A$} Let us now derive the Hilbert-Poincar\'e series 
\be P_A(q,t) := \sum_{p,s \geq 0} \text{dim}(A^{(s,p)}) q^s t^p, \ee 
of the Poisson vertex algebra $A$. The claim is that 
\begin{align} \label{hpclaim}
P_A(q,t) = \sum_{n =0}^{\infty} \frac{(-tq ;q)_n}{(q; q)_n} t^{2n} q^{n(n+1)},
\end{align}
where $(a;q)_n$ is the standard Pochhammer symbol \be (a;q)_n = \prod_{k=0}^{n-1} (1-a q^k).\ee

In order to prove the Hilbert-Poincar\'e series of $A$ is as claimed, we study its graded dual $A^{\vee}$. We first form generating functions of the fields 
\begin{align}
    X(z) = \sum_{k=0}^{\infty} \del^k X \frac{z^k}{k!} , \,\,\,\,\,\,\, Y(z) = \sum_{k=0}^{\infty} \del^k Y \frac{z^k}{k!}.
\end{align} Introduce an anti-commuting variable $\theta$ carrying $s(\theta) = g(\theta)=-1$, and consider the ghost number 2, spin 2 superfield \be \mathcal{K}(z,\theta) = X(z) + \theta Y(z) \ee A degree $n$ (where degree just refers to the the graded polynomial degree, not to be confused with spin and ghost number degrees) element in the differential supercommutative algebra $V$ can be obtained from coefficients of 
\be \mathcal{K}(z_1, \theta_1) \dots \mathcal{K}(z_n, \theta_n).\ee The coefficient of 
\be  z_1^{k_1} \dots z_n^{k_n}  \theta_{i_1} \dots \theta_{i_p} \ee gives us the monomial \be \frac{1}{k_1! \dots k_n!}\del^{k_1} Z_{a_1}\dots \del^{k_n} Z_{a_n} \ee where 
\begin{align} Z_{a_i} &= Y  \text{ for } i \in \{i_1, \dots, i_p \}, \\ Z_{a_i} &= X, \text{otherwise}. \end{align} An element in the dual $\eta \in V^{\vee}$ thus gives rise to 
\be \omega(z_1, \dots, z_n, \theta_1, \dots, \theta_n) = \eta \big( \mathcal{K}(z_1, \theta_1) \dots \mathcal{K}(z_n, \theta_n) \big)\ee where $\omega(z_1, \dots, z_n, \theta_1, \dots, \theta_n)$ is a polynomial in the commuting variables $(z_1, \dots, z_n)$ and anti-commuting variables $(\theta_1, \dots, \theta_n)$. Now note that since $X(z)$ is commutative and $Y(z)$ is anti-commutative, the superfield $\mathcal{K}(z,\theta)$ is commutative \be \mathcal{K}(z_1,\theta_1) \mathcal{K}(z_2, \theta_2) = \mathcal{K}(z_2, \theta_2) \mathcal{K}(z_1, \theta_1). \ee Therefore $\omega(z_1, \dots, z_n, \theta_1, \dots, \theta_n)$ must be invariant under simultaneous interchanges of $(z_i, \theta_i)$ and $(z_j, \theta_j)$, 
\be \omega( \dots, z_i, \dots, z_j,  \dots, \theta_i, \dots, \theta_j, \dots) = \omega(\dots , z_j, \dots, z_i, \dots, \theta_j, \dots, \theta_i, \dots). \ee By thinking of $\theta_i$ and the one form $dz_i$, this is nothing but the algebra of polynomial differential forms on $\mathbb{C}^n$ that are invariant under the action of the symmetric group $S_n$ acting on $\mathbb{C}^n$ by permuting the coordinates. 

Thus we see that the dual space of the degree $n$ part of $V$ corresponds to the ring of $S_n$-invariant polynomial differential forms on $\mathbb{C}^n$. We now work out when an element $\eta \in V^{\vee}$ descends to a dual of the quotient $V/I$. Note that the generators of the differential ideal of relations simply corresponds to the coefficients of the fields \be X(z)^2, \,\,\, X(z) Y(z), \,\,\, X(z) \frac{\del}{ \del z} Y(z), \,\,\,\,\, Y(z) \frac{\del}{\del z} Y(z).\ee These can be neatly packaged in terms of the superfield $\mathcal{K}(z,\theta)$: the generators simply follow from the two coefficients of the superfields
\be \mathcal{K}(z, \theta_1) \mathcal{K}(z, \theta_2), \,\,\,\,\,\, \mathcal{K}(z, \theta_1) \frac{\del}{\del z} \mathcal{K}(z, \theta_2). \ee Indeed expanding out the first gives rise to $X(z)^2$ and $X(z) Y(z)$, whereas expanding the second gives rise to the new independent generators $X(z) \frac{\del}{\del z} Y(z)$ and $Y(z) \frac{\del}{\del z} Y(z)$. Thus for $\eta$ to descend to $A$, we must have \be \eta \big(\mathcal{K}(z_1, \theta_1) \dots \mathcal{K}(z_n, \theta_n) \big) |_{z_i = z_j} = 0, \,\,\,\,\,\, i \neq j,\ee  which implies $\omega(z_1, \dots, z_n, \theta_1, \dots, \theta_n)$ is divisible by $z_i - z_j$ for each $i \neq j$. In other words, $\omega$ takes the form \be \omega = \Delta(z_1, \dots, z_n) \chi\ee where 
\be \Delta(z_1, \dots, z_n) = \prod_{1 \leq i < j \leq n} (z_i - z_j)\ee is the standard Vandermonde factor. On the other hand, by the second relation we must also have \be \big(\frac{\del}{\del z_i} \eta \big( \mathcal{K}(z_1, \theta_1) \dots \mathcal{K}(z_n, \theta_n) \big) \Big)|_{z_i = z_j} = 0 ,\ee which means $\del_{z_i} \omega$ must also be divisible by $(z_i - z_j)$ for each $i \neq j$. Together these imply that $\omega$ must have the square of the Vandermonde determinant as a factor: 
\begin{align}
    \omega = \big(\Delta(z_1, \dots, z_n) \big)^2 \mu
\end{align} Now since the square of the Vandermonde is an $S_n$-invariant zero-form, the factor $\mu$ must also be an $S_n$-invariant differential form. We thus see that an element $\eta$ descends to the dual only if it is an invariant differential form carrying $\Delta^2$ as a factor. 

Conversely, any polynomial $S_n$-invariant differential form which is divisible by the square of the Vandermonde factor vanishes at $z_i = z_j$, and also has the property that $\frac{\del}{\del z_i}$ vanishes at evaluating any of the $z_{j}$ for $j \neq i$ at $z_i = z_j$. Therefore one has the isomorphism 
\be A^{\vee}_n \cong \Delta_n^2 \Big(\mathbb{C}[z_1, \dots, z_n] \otimes \Lambda[dz_1, \dots, dz_n] \Big)^{S_n}. \ee

The final fact we need to proceed with the computation of the Hilbert-Poincar\'e series is the description of the ring of polynomial $S_n$-invariant differential forms in $n$ variables $(z_1, \dots, z_n)$. This is well-known as a special case of Solomon's theorem. Letting 
\begin{align}
e_k = \sum_{1 \leq i_1 < \dots <i_k \leq n} z_{i_1} \dots  z_{i_k},
\end{align} be the $k$th elementary symmetric polynomial for $k \in \{1,\dots, n\}$, and letting \begin{align} d e_k = \sum_{i=1}^n \frac{\del e_k}{\del z_i} \text{d}z_i, \end{align}
be its differential, we have an isomorphism of graded algebras
\begin{align}
    \big(\mathbb{C}[z_1, \dots, z_n] \otimes \Lambda[dz_1, \dots, dz_n] \big)^{S_n} \cong \mathbb{C}[e_1, \dots, e_n] \otimes \Lambda[de_1, \dots, de_n].
\end{align} Since both $z_i$ and $dz_i$ are assigned to have $-1$, the spins of both $e_k$ and $de_k$ is $-k$, whereas the ghost numbers are $(0,-1)$ respectively. Thus the space of $S_n$-invariant polynomial differential forms has the Hilbert-Poincar\'e series
\begin{align}
    \prod_{k=1}^n \frac{1+ t^{-1} q^{-k}}{1-q^{-k}}.
\end{align}
Next, we must remember that an element $\eta$ carries base spin $-2n$ and base cohomological degree $-2n$ coming from the spin and ghost number of the $n$ insertions of the superfield $\mathcal{K}.$ Finally, the Vandermonde squared factor carries spin $-n(n-1)$, giving rise to the overall factor $q^{-n(n+1)} t^{-2n}$. Combining these factors, and dualizing gives us \be q^{n(n+1)} t^{2n} \prod_{k=1}^n \frac{1+t q^k}{1-q^k},\ee the graded dimensions of the bigraded pieces of the degree $n$ part of $A$. Summing over all $n$ thus readily gives us \begin{align}
P_A(q,t) = \sum_{n =0}^{\infty} \frac{(-tq ;q)_n}{(q; q)_n} t^{2n} q^{n(n+1)},
\end{align} as desired.

Finally we mention a refinement by a ``$B$-number" that will be used later. We assign $X$ a $B$-number of $+1$ and $Y$ a $B$-number of $+3$,
\begin{align}
B(\del^k X) = 1, \,\,\, B(\del^k Y) = 3,
\end{align} so that a superfield 
\begin{align} 
\mathcal{K}(z,\theta) = X(z) + \theta\,Y(z), 
\end{align} has $B$-number $+1$ provided we assign the commuting variable $z$ a vanishing $B$-number, and the anti-commuting variable $\theta$ a $B$-number of $-2$, 
\begin{align}
    B(z) = 0,\,\,\,\,\,\,\,\,\,\,\,B(\theta) = -2.
\end{align}
Letting the variable $y$ keep track of the $B$-number, the ring of $S_n$-invariant differential forms thus now has the $B$-refined character
\begin{align}
    \prod_{k=1}^n \frac{1+t^{-1} y^{-2} q^{-k}}{1-q^{-k}}.
\end{align} Dualizing and shifting with the base spin and Vandermonde-squared factor as before, yields the $B$-refined Hilbert-Poincar\'e series of $A$: \be P_A(q,t,y) = \sum_{n = 0}^{\infty} q^{n(n+1)} t^{2n}y^{n} \frac{(-t q y^2 ; q)_n}{(q; q)_n}.\ee 

Let us now connect the Poisson vertex algebra $A$ with the algebra of holomorphic-topological observables of $\mathcal{N}=2$, $SU(2)$ gauge theory.

\section{Holomorphic-Topological Observables}  \label{sw} 
\subsection{Holomorphic-Topological Twist}
Recall in four-dimensional $\mathcal{N}=2$ supersymmetry we have supercharges $Q^A_{\alpha}$ and $\ov{Q}^A_{\dot{\alpha}}$ where $A$ is an index for the $SU(2)_R$ symmetry, and $\alpha, \dot{\alpha}$ are Lorentz indices. The supersymmetry algebra as usual is 
\begin{align}
    \{Q^A_{\alpha}, \ov{Q}_{\dot{\beta} }^B \} = 2 \epsilon^{AB} \sigma^{\mu}_{\alpha \dot{\beta}} P_{\mu}. 
\end{align} There are also central charges present as usual, 
\begin{align}
    \{Q^A_{\alpha}, Q^B_{\beta} \} = 2 \ov{Z} \epsilon^{AB} \epsilon_{\alpha \beta}
\end{align} along with the Hermitian conjugate of this equation. The supercharges are subject to the constraint \be (Q^A_{\alpha})^{\dagger} = Q_{\dot{\alpha} A} :=  \epsilon_{AB} \ov{Q}_{ \dot{\alpha}  }^B.\ee The $U(1)_r$ symmetry is also important in what follows. One has that $Q^A_{\alpha}$ has $r$-charge $+1$ and $\ov{Q}^A_{\dot{\alpha}}$ has $r$-charge $-1$.

In the holomorphic-topological twist, the theory we obtain fundamentally breaks the $SO(4)$ Lorentz symmetry (since by definition one picks out ``holomorphic" and "topological" directions and treats those differently). Accordingly, one only considers the rotation subgroup $SO(2) \times SO(2) \subset SO(4)$ defined as follows. Writing the coordinates on $\mathbb{R}^4$  as $(x^1, x^2, x^3, x^4)$ the first $SO(2)$ is the rotation generator in the $(12)$-plane whereas the second $SO(2)$ factor is the rotation generator in the $(34)$-plane. Letting $J_{12}$ and $J_{34}$ be the infinitesimal rotation generator in each of these planes, the $SO(2)_{12}$ and $SO(2)_{34}$ charges of the supercharges can be read off as follows. 

\begin{table}[h]
\centering
\renewcommand{\arraystretch}{1.25}
\begin{tabular}{c|cc}
Supercharge & $q_{12}$ (SO(2)$_{12}$) & $q_{34}$ (SO(2)$_{34}$) \\
\hline
$Q^{A}_{1}$              & $+\tfrac12$ & $+\tfrac12$ \\
$Q^{A}_{2}$              & $-\tfrac12$ & $-\tfrac12$ \\
$\bar Q_{\dot 1}^A$     & $-\tfrac12$ & $\tfrac12$ \\
$\bar Q_{\dot 2}^A$     & $+\tfrac12$ & $-\tfrac12$ \\
\end{tabular}
\caption{Charges of 4d $\mathcal N=2$ supercharges under rotations in the $(12)$ and $(34)$ planes (Euclidean signature). The $SU(2)_R$ index $A$ is a spectator for these spacetime charges.}
\end{table}

The holomorphic-topological twist corresponds to mixing each of these $SO(2)$'s with an R-charge. Letting $I_3$ be the Cartan of the $SU(2)_R$ and $r$ be the generator of the $U(1)_r$, we consider the ``twisted" Lorentz subgroup (isomorphic to $SO(2) \times SO(2)$) to be defined by the generators 
\begin{align}
   J_{12}' &= J_{12} + I_3, \label{ts1}  \\  J_{34}' &= J_{34} + \frac{1}{2}r. \label{ts2}
\end{align}
This combination picks out two supercharges that are scalars: namely $Q^+_{2}$ and $\ov{Q}^+_{\dot{1}}:$ \begin{align}
    [J'_{12}, Q^+_2] &= [J'_{34}, Q^+_2] = 0, \\ 
    [J'_{12}, \ov{Q}^+_{\dot{1}}] &= [J'_{34}. \ov{Q}^+_{\dot{1}}] = 0.
\end{align} The \textit{holomorphic-topological} supercharge is then defined to be a linear combination of the above \begin{align}
    Q_{\text{HT}} = \ov{Q}^+_{\dot{1}} + Q^+_{2} . 
\end{align} With this twisted spin, it is a scalar, square-zero operator \be (Q_{\text{HT}})^2 = 0, \ee that moreover renders the translations $\del_3, \del_4$ and $ \del_{\bar{z}} := \del_1 +i \del_2$ as null-homotopic: 
\begin{align}
    \{Q_{\text{HT}}, Q^-_{1} \} &= -P_w + \ov{Z}, \\
    \{Q_{\text{HT}}, \ov{Q}^-_{\dot{2}} \} &= -P_{\ov{w}} + Z, \\
    \{Q_{\text{HT}}, \ov{Q}^-_{\dot{1}} \} &= P_{\bar{z}},
\end{align} where $w = x^3 + i x^4$ and $z = x^1 + i x^2.$

In the ``taxonomy" of \cite{Elliott:2020ecf}, $Q_{\text{HT}}$ is a specific representative of the (unique) orbit of the nilpotent supercharges that lie in the rank $(1,1)$ twist \footnote{In particular there is only a single holomorphic-topological twist allowed by the four-dimensional $\mathcal{N}=2$ algebra, up to equivalence. }.

\begin{remark} $Q_{\text{HT}}$ lies in the BPS subalgebra. In particular, it is one of the four generators that annihilate a BPS state. Specifically in the notation of \cite{moorelectures} where the BPS supercharges are denoted as \be \mathcal{R}^A_{\alpha} = \xi^{-1} Q^A_{\alpha} + \xi \sigma^0_{\alpha \dot{\beta}} \ov{Q}^{A \dot{\beta}},\ee one has $Q_{\text{HT}} = \mathcal{R}^+_2,$ with $\xi = 1$.  We expect this observation to be an important starting point for a first-principles derivation of the various UV-IR relations observed in the literature \cite{Cecotti:2010fi, Cordova:2015nma, Cecotti:2015lab,Cordova:2016uwk, Andrews:2025tko}. We hope to return to this point in future work. \end{remark}

\subsection{Holomorphic-Topological BF Theory}
Let us now specialize to pure $\mathcal{N}=2$ gauge theory, namely the four-dimensional $\mathcal{N}=2$ gauge theory with a vector multiplet in the adjoint representation of a gauge group $G$. In order to determine the space of local observables in $Q_{\text{HT}}$-cohomology (at leading order in perturbation theory) one can simply take the standard vector multiplet fields \be (A_{\mu}, \phi, \bar{\phi},  \lambda^A_{\alpha}, \ov{\lambda}^A_{\dot{\alpha}}, D^I),\ee and specialize the four-dimensional $\mathcal{N}=2$ supersymmetry transformations of the vector multiplet to the linear combination picked out by the holomorphic-topological supercharge $Q_{\text{HT}},$ renaming the fields according to the twisted spins defined by \eqref{ts1}, \eqref{ts2}. While this is straightforward\footnote{and can indeed be found, for instance in Equation 3.3 of \cite{Jeong:2019pzg}.}, a more elegant formulation can be obtained by using the description of twisted supersymmetric Yang-Mills theories in \cite{Elliott:2020ecf}.

Theorem $10.7$ of \cite{Elliott:2020ecf} (see also \textbf{Claim} in Section 0.16 of \cite{Costello:2013zra}) says that the holomorphic-topological twist of the four-dimensional $\mathcal{N}=2$ theory of a vector multiplet for a gauge group $G$ is perturbatively equivalent to the four-dimensional holomorphic-topological BF theory for the Lie algebra $\mathfrak{g} = \text{Lie}(G).$

Since the latter is best described using the BV formalism, let's provide a brief recall. In the BV formalism a (classical) field theory is described by a supermanifold $\mathcal{M}$ \footnote{Roughly, the full space of fields, ghosts, anti-fields.}, equipped with an odd symplectic form $\omega$, and a homological vector field $Q$ \cite{Alexandrov:1995kv}: an odd vector field $Q$ that preserves the odd symplectic form $\mathcal{L}_Q \omega = 0$ and moreover satisfies \be [Q,Q] = 0. \ee Equivalently, the odd symplectic form defines the BV bracket $\{ \, , \, \}_{\text{BV}}$ (a shifted Poisson bracket) and the homological vector field $Q$ defines the moment map $S$ that satisfies the classical master equation \be \{S,S \}_{\text{BV}} = 0.\ee 

In the holomorphic-topological $BF$ theory formulated on $\mathbb{R}^2 \times \mathbb{C}$, $\mathcal{M}$ consists of \be \mathcal{M} = \Pi \big( \Omega_{\text{HT}}(\mathbb{R}^2 \times \mathbb{C}) \otimes \mathfrak{g} \big) \times \Pi \big( \text{d}z \,\Omega_{\text{HT}}(\mathbb{R}^2 \times \mathbb{C}) \otimes \mathfrak{g}^{\vee}\big) \ee where 
\begin{align}
    \Omega_{\text{HT}}(\mathbb{R}^2 \times \mathbb{C}) = \Omega_{\text{dR}}(\mathbb{R}^2) \otimes \Omega^{(0,*)}_{\text{Dolb}}(\mathbb{C}), 
\end{align} is the space of mixed forms on $\mathbb{R}^2 \times \mathbb{C}$ spanned by $\text{d}x, \text{d}y, \text{d} \bar{z}$ (in particular no $\text{d}z$). Thus the field space consists of a field
\begin{align}
\mathbf{b} \in \Pi \big( \text{d}z \,\Omega_{\text{HT}}(\mathbb{R}^2 \times \mathbb{C}) \otimes \mathfrak{g}^{\vee} \big) 
\end{align} which is a co-adjoint valued mixed-form of the type
\begin{align}
    \mathbf{b} = \text{d}z \big(b^{(0)} + b^{(1)} + b^{(2)} + b^{(3)} \big)
\end{align} along with a field  
\begin{align}
    \mathbf{c} \in \Pi \big( \Omega_{\text{HT}}(\mathbb{R}^2 \times \mathbb{C}) \otimes \mathfrak{g} \big)
\end{align} which is an adjoint-valued mixed form of the type
\begin{align}
    \mathbf{c} = c^{(0)} + c^{(1)} + c^{(2)} + c^{(3)}.
\end{align} Because $\mathcal{M}$ is a supermanifold, we have to specify the parity. For both the $\mathbf{b}$ and $\mathbf{c}$ fields we take the parity $P$ of the $p$-form component to be given by $(-1)^{p+1}$:
\begin{align}
    P(\text{d}z \,b^{(p)}) = P( c^{(p)}) = (-1)^{p+1}.
\end{align} Thus, in particular the lowest components of both $\mathbf{b}$ and $\mathbf{c}$ are anti-commuting. In addition, we assign a $\mathbb{Z}$-valued gradation by the ghost number, denoted $\text{gh}$ to the fields by setting 
\begin{align}
    \text{gh}(\text{d}z \, b^{(p)}) = \text{gh}(c^{(p)}) = 1-p.
\end{align}

The odd symplectic form on $\mathcal{M}$ is given by 
\begin{align}
    \omega = \int_{\mathbb{R}^2 \times \mathbb{C}}  \delta \mathbf{b}_a  \wedge \delta \mathbf{c}^a. 
\end{align}  
Finally, to specify the homological vector field $Q$, we first consider the odd vector field on $\mathcal{M}$ given by the mixed deRham-Dolbeault differential \be d_{\text{HT}} = d_{\mathbb{R}^2} + \ov{\del}_{\mathbb{C}} = \text{d}x \frac{\del}{\del x} + \text{d}y \frac{\del}{\del y } + \text{d} \bar{z} \frac{\del }{\del \bar{z}}.\ee The homological vector field $Q$ that defines holomorphic-topological BF theory is given by the vector field $Q$ defined via
\begin{align}
Q \mathbf{b}_a &= \text{d}_{\text{HT}} \mathbf{b}_a  + f_{ab}^c \,\mathbf{c}^b \, \mathbf{b}_c, \\ Q \mathbf{c}^a &= \text{d}_{\text{HT}} \mathbf{c}^a + \frac{1}{2} f^a_{bc} \, \mathbf{c}^b \, \mathbf{c}^c. 
\end{align} Equivalently, the BV action of holomorphic-topological BF theory is given by \be S(\mathbf{b}, \mathbf{c}) = \int_{\mathbb{R}^2 \times \mathbb{C}} \mathbf{b} \, \text{d}_{\text{HT}} \,\mathbf{c} + \frac{1}{2} \mathbf{b}[\mathbf{c}, \mathbf{c}] .\ee
$S$ satisfies the classical master equation \be \{ S,S \}_{\text{BV}} = 0\ee (equivalently $[Q,Q]=0$) due to the Jacobi identity of $\mathfrak{g}$. 

\subsection{BRST Reduction of Classical $bc$ System from BF Theory}

The perturbative equivalence between the holomorphic-topological BF theory and the holomorphic-topological twist of the theory of pure vector multiplet guarantees in particular that the space of point-like, zero-form observables in the cohomology of the BV differential $Q_{\text{BV}}$ is isomorphic to the cohomology of local operators in the cohomology of $Q_{\text{HT}}$. Let us then study the space of local operators in the BV formulation.

Let's first recover the by now familiar statement that is often found in the literature on (Poisson) vertex algebras of four-dimensional $\mathcal{N}=2$ theories. The theory of a pure vector multiplet is supposed to be related to the classical BRST reduction of a $bc$-ghost system with $b$ valued in $\mathfrak{g}^{\vee}$ and $c$ valued in $\mathfrak{g}$ \cite{Beem:2013sza, Oh:2019mcg}. This is straightforward to derive from the BV formulation of the holomorphic-topological BF theory we formulated in the previous section. 

In order to work out the space of local observables in the neighborhood of a point in $\mathbb{R}^2 \times \mathbb{C}$, we consider the infinite jet space of graded-commutative polynomials on $b^{(k)}, c^{(k)}$, namely the exterior algebra in 
\begin{align} \del_x^{k} \del_y^l \del_{z}^m \del_{\bar{z}}^n b^{(k)}, \,\,\,\, \del_x^{k} \del_y^l \del_{z}^m \del_{\bar{z}}^n c^{(k)} 
\end{align} where we evaluate the derivative at the point we specified. 
We then look at the action of the BV differential acting on these jets. We recall that 
\begin{align}
Q_{\text{BV}} \mathbf{b}_a &= \text{d}_{\text{HT}} \mathbf{b}_a  + f_{ab}^c \,\mathbf{c}^b \, \mathbf{b}_c, \\ Q_{\text{BV}} \mathbf{c}^a &= \text{d}_{\text{HT}} \mathbf{c}^a + \frac{1}{2} f^a_{bc} \, \mathbf{c}^b \, \mathbf{c}^c, 
\end{align} which is extended to act on the infinite jet algebra by requiring $Q$ to commute with derivatives, and requiring it to be an odd derivation. 
Since the differential $Q_{\text{BV}}$ splits as a sum of the action of $d_{\text{HT}}$ (coming from the ``free" part of the BV action), and the interacting part coming from the gauge algebra, one can proceed sequentially. The cohomology of $d_{\text{HT}}$ is determined from the formal/algebraic Poincar\'e Lemma. This says that the cohomology is simply concentrated in form degree zero, and is given by the \textit{holomorphic} jets of the zero-form components $b=b^{(0)}$ and $c = c^{(0)}$. Letting $\del = \del_z$, So the $\text{d}_{\text{HT}}$-cohomology is given the exterior algebra 
\be H^*_{\text{d}_{\text{HT}}} = \mathcal{C}_{bc}(\mathfrak{g}) := \Lambda(c, \del c, \del^2 c, \dots, b, \del b, \del^2 b, \dots).  \ee 
The remaining, ``interacting" part of the differential then simply acts on these via the usual classical BRST differential
\begin{align} \label{brst}
    Q b_a &= f_{ab}^c \, c^b \,b_c, \\
    Q c^a &= \frac{1}{2} f^a_{bc} \,c^b c^c ,  
\end{align} extended to act on holomorphic jets as before.

Recall also the gradings on $\mathcal{C}_{bc}(\mathfrak{g})$: the ghost numbers of $b$ and $c$ were taken to be $+1$, whereas the spin of $b$ was $+1$ and spin of $c$ is zero. The derivation $\del$ keeps the ghost number unchanged, whereas it raises the spin by $+1$:
\begin{align} 
s(\del^{k} b) &= s(\del^{k+1} c) = k+1, \\
\text{gh}(\del^k b) &= \text{gh}(\del^k c) = 1,
\end{align} so that 
\begin{align}
    \mathcal{C}_{bc}(\mathfrak{g}) = \bigoplus_{s, \text{gh}} \mathcal{C}^{s, \text{gh}}
\end{align}The differential $Q$ indeed raises the ghost number by $+1$ and preserves spin,
\begin{align}
    Q: \mathcal{C}^{s, \text{gh}} \to \mathcal{C}^{s, \text{gh}+1}.
\end{align}

An additional grading that is useful to introduce (as in Section \ref{PVA}) is the $B$-number. As the notation suggests, we assign the field $c$ a vanishing $B$-number and the field $b$ a $B$-number of $+1$:
\begin{align}
    B(\del^k c) = 0, \,\,\,\,\,\,\,\,\,\,\,\,\,B(\del^k b) = 1. 
\end{align} The BRST differential $Q$ preserves the $B$-number.

So far $\mathcal{C}_{bc}(\mathfrak{g})$ is just a differential-graded (anti)-commutative algebra. There is also however a standard (shifted) Poisson vertex algebra structure on it. Since $\mathcal{C}_{bc}(\mathfrak{g})$ is a free supercommutative algebra, specifying a $\lambda$-bracket just requires us to specify the $\lambda$-bracket of generators. A straightforward tree-level calculation \cite{Oh:2019mcg} gives the standard $\lambda$-bracket
\begin{align} \label{bc}
    \{b_a \,_{\lambda} \, c^b\} = \delta_a^{\,\,b},
\end{align} with the $\lambda$-bracket of $b$ with itself and $c$ with itself vanishing. This indeed gives the $\lambda$-bracket ghost number $-2$ and spin $-1$ as discussed earlier in Section 2. 

The differential $Q$ is a derivation of the $\lambda$-bracket namely 
\begin{align}
    Q\{ x\,_{\lambda} \,y\} = \{Q x \,_{\lambda} \,y\} + (-1)^{p(x)} \{x \,_{\lambda}\, Qy \},  
\end{align} as can be readily checked on the defining $\lambda$-bracket \eqref{bc}.

Thus we've recovered the classical $bc$-ghost system $\mathcal{C}_{bc}(\mathfrak{g})$ as a differential-graded Poisson vertex algebra 
\begin{align}
(\mathcal{C}_{bc}(\mathfrak{g}), \cdot, \del, \{ \cdot \,_{\lambda} \, \cdot \} , Q), 
\end{align}
from the BV cohomology of the four-dimensional holomorphic-topological theory. However, $\mathcal{C}_{bc}(\mathfrak{g})$ (and even its $Q$-cohomology) is \textit{not} the correct answer for the space of physical local observables of the holomorphic-topological twist of a $\mathfrak{g}$-valued $\mathcal{N}=2$ vector multiplet. The reason is that we have not yet implemented the effect of ``constant" gauge transformations. In the BV/BRST formalism, the ghost zero-mode, namely $c$ without any holomorphic derivatives is what is responsible for this. The proper way to deal with the effect of these gauge transformations is to pass to $\textit{basic}$ elements with respect to $\mathfrak{g}$.

\subsection{Basic Invariants and $\text{Obs}_{\text{HT}}(\mathfrak{g})$.}
We now apply a general construction, which goes as follows. Suppose $V$ is a differential-graded Poisson vertex algebra, and $T$ is a Lie algebra with a basis $\{t_\alpha\}.$ Suppose there is an action of $T$ on $V$ by derivations of the (super-commutative) product, derivations of the $\lambda$-bracket, and commuting with $Q$ and $\del$. Explicitly, letting $L_\alpha : V \to V$ denote the action of $t_{\alpha}$ on $V$, we require
\begin{align}
    [L_{\alpha}, L_{\beta}] &= f_{\alpha \beta}^{\gamma} L_{\gamma}, \\ 
    L_{\alpha}(x \cdot y) &= L_{\alpha}(x) \cdot y + x \cdot L_{\alpha} (y), \\
    L_{\alpha} \{ x \,_{\lambda} \, y \} &= \{L_{\alpha}x \,_{\lambda} \,y \} + \{ x \,_{\lambda} \, L_{\alpha} y \} , \\ [\del, L_{\alpha}] &= [ Q, L_{\alpha}] = 0. 
\end{align} Moreover, suppose that we also have ``contraction" operators for each $t_{\alpha}$ i.e maps $\iota_{\alpha} : V \to V$ of degree $-1$, which also commute with $\del$ are derivations of $\cdot$ and the $\lambda$-bracket, and which mutually commute:
\begin{align}
\iota_{\alpha}(x \cdot y) &= \iota_{\alpha}(x) \cdot y + (-1)^{p(x)}x \cdot \iota_{\alpha} (y), \\
        \iota_{\alpha} \{ x \,_{\lambda} \, y \} &= \{\iota_{\alpha}x \,_{\lambda} \,y \} + (-1)^{p(x)}\{ x \,_{\lambda} \, \iota_{\alpha} y \}, \\
        [\iota_{\alpha}, \iota_{\beta}] &= [\del, \iota_{\alpha}] = 0.
\end{align} The contraction operators $\iota_{\alpha}$ and $L_{\alpha}$ are further assumed to satisfy $T$-equivariance,
\begin{align} 
    [L_{\alpha}, \iota_{\beta}] &= f_{\alpha \beta}^{\gamma} \,\iota_{\gamma},
\end{align} and the Cartan formula
\begin{align}
   L_{\alpha} &=  \{ Q, \iota_{\alpha} \}.
\end{align} If these conditions are satisfied, we say that $V$ is a $T$-differential Poisson vertex algebra.

If $V$ is a $T$-differential Poisson vertex algebra, we may define the $T$-basic elements of $V$ as the joint kernel of $L_{\alpha}$ and $i_{\alpha}$ inside $V$:
\begin{align}
    V_{\text{bas}}^{T} = \{x \in V | L_{\alpha}( x) = \iota_{\alpha}(x) = 0, \text{ for all } \alpha \}. 
\end{align} The standard terminology is that elements in the kernel of the contraction operators $\iota_{\alpha}$ are said to be \textit{horizontal}, whereas those in the kernel of the $L_{\alpha}$ operators are said to be \textit{invariant}. A basic element is both horizontal and invariant.  The advantage of having this Cartan calculus is that $V_{\text{bas}}^T$ then automatically acquires the structure of a differential-graded Poisson vertex algebra. Indeed both $Q$ and  $\del$ act on the $T$-basic elements, since $Q$ and $\del$ both commute with $L_{\alpha}$, and  
\begin{align}
    \iota_{\alpha}(Q x) = L_{\alpha} x -Q(i_{\alpha} (x) ) = 0, 
\end{align} and the basic elements are also closed under the supercommutative product and $\lambda$-bracket since both $L_{\alpha}$ and $\iota_{\alpha}$ act as derivations of both. Thus $V^{T}_{\text{bas}}$ is a differential-graded Poisson vertex algebra. 

Let us now apply this to the case we're interested in. Consider the dg-PVA $\mathcal{C}_{bc}(\mathfrak{g})$. We claim that $\mathcal{C}_{bc}(\mathfrak{g})$ is a $\mathfrak{g}$-differential Poisson vertex algebra. The operators $L_a$ simply act in the obvious way
\begin{align}
    L_{a}(b_b) &= f_{ba}^c \,b_c, \\
    L_{a}(c^b) &=  f_{ac}^b \, c^c,
\end{align} requiring to commute with $\del$, extended to act as derivations. The contraction $\iota_a$ on the other hand is defined via 
\begin{align}
    \iota_a = \frac{\del}{\del c^a},
\end{align} said differently, 
\begin{align}
    \iota_{a}(\del^k c^b) = \delta^k_0 \delta^b_a, \,\,\,\, \iota_a(\del^k b) = 0. 
\end{align} It's then straightforward to check the Cartan formula and $\mathfrak{g}$-equivariance holds. 

We can then pass to $\mathfrak{g}$-basic elements, defining 
\begin{align}
    \mathcal{C}_{\text{HT}}(\mathfrak{g}) :=  \big(\mathcal{C}_{bc}(\mathfrak{g}) \big)^{\mathfrak{g}}_{\text{bas}},
\end{align} with its inherited dg PVA structure. Finally, the physical space of holomorphic-topological observables is the cohomology Poisson vertex algebra,
\begin{align}
    \text{Obs}_{\text{HT}}(\mathfrak{g}) = H^*\big(\mathcal{C}_{\text{HT}}(\mathfrak{g}) \big).
\end{align}

It is now straightforward to identify the $\mathfrak{g}$-basic elements. Any exterior element built from $\del c, \del^2c , b, \del b, \del^2 b, \dots$  is killed by the contraction operators $\{i_a \}$ and vice-versa. On the other hand, being in the kernel of $\{L_a\}$ simply imposes the condition that all words we build must be invariant under the natural $\mathfrak{g}$-action. Thus we have as a graded vector space 
\be \mathcal{C}_{\text{HT}}(\mathfrak{g}) =  \big( \Lambda(\del c, \del^2 c, \dots, b, \del b, \del^2 b , \dots) \big)^{\mathfrak{g}}.\ee
The differential that acts on $Q$ is simply the Chevalley differential as before. Note that as discussed before, as a consequence of the Cartan formula
\begin{align}
    L_a = Q \iota_a + \iota_a Q,
\end{align} the Chevalley differential preserves basic elements. In particular, when acting with the Chevalley differential $Q$ on any basic element, even though a priori, one could get contributions from terms involving the bare ghost $c$, the Cartan formula guarantees that any dependence on those terms ultimately drops out when $Q$ acts on basic elements.
Therefore in the basic complex, we can define $Q$ by introducing a field $\widetilde{c}(z)$ with the constant mode omitted. Letting
\begin{align}
    b_a(z) = \sum_{n \geq 0} \del^n b_a \frac{z^n}{n!}, \,\,\,\,\,\, \tilde{c}^a(z) = \sum_{n \geq 1} \del^n c^a \frac{z^n}{n!}, 
\end{align} the differential then simply acts on basic elements as
\begin{align}
Q b_a(z) &= f_{ab}^c\, \tilde{c}^b(z) \, b_c(z), \\ Q \tilde{c}^a(z) &= \frac{1}{2} f^a_{bc} \, \tilde{c}^b(z)\, \tilde{c}^c(z),    
\end{align} We thus obtain a differential-graded PVA given explicitly as 
\begin{align}
    \mathcal{C}_{\text{HT}}(\mathfrak{g}) =  \Big(  \Lambda(\del c, \del^2 c, \dots, b, \del b, \del^2 b , \dots) ^{\mathfrak{g}}  , \del, Q, \cdot, \{\cdot \,_{\lambda} \,\cdot\} \ \Big).
\end{align} The algebra of holomorphic-topological observables is thus
\begin{align}
    \text{Obs}_{\text{HT}}(\mathfrak{g}) = H^*_Q\big(\mathcal{C}_{\text{HT}}(\mathfrak{g}) \big).
\end{align}

Forgetting the Poisson vertex algebra structure, the graded vector space $\text{Obs}_{\text{HT}}(\mathfrak{g})$ we have arrived at can be expressed cleanly in terms of relative Lie algebra cohomology. Letting $\mathfrak{h}$ be a Lie algebra, $ \mathfrak{f} \subset \mathfrak{h}$ a Lie subalgebra and $M$ an $\mathfrak{h}$-module, the Lie algebra cohomology of $\mathfrak{h}$ relative to $\mathfrak{f}$ with values in $M$, denoted as \begin{align}
H^*_{\text{CE}}(\mathfrak{h}, \mathfrak{f} \,; M)  ,  
\end{align} is defined precisely as the cohomology of the complex of $\mathfrak{f}$-basic elements in the standard Chevalley complex 
\begin{align}
    \text{CE}(\mathfrak{h};M) = \text{Hom}(\Lambda^* \mathfrak{h}, M). 
\end{align} The $\mathfrak{f}$-basicness condition is defined precisely as before, since each element $x \in \mathfrak{f}$ defines a pair of operators $(\iota_x, L_x)$ satisfying the Cartan formula \begin{align}
L_x = \{Q_{\text{CE}}, i_x\}.
\end{align}
Applying the construction with \be \mathfrak{h} = \mathfrak{g}[[z]], \,\,\,\,\, \mathfrak{f} = \mathfrak{g}, \,\,\,\,\, M = \Lambda\big( \, (\mathfrak{g}[[z]] \,\text{d}z)^{\vee}\big)\ee recovers \be \text{Obs}_{\text{HT}}(\mathfrak{g}) = H^*\big( \mathfrak{g}[[z]], \mathfrak{g} ; \, \Lambda(   (\mathfrak{g}[[z]] \,\text{d}z)^{\vee} ) \,\big) .\ee
An equivalent way to rewrite this in terms of relative Lie algebra cohomology uses the (even) Lie algebra 
\begin{align}
    \mathfrak{g}[z, \varepsilon] = \mathfrak{g}[[z]] \ltimes \varepsilon \, \mathfrak{g}[[z]]
\end{align} where $\varepsilon$ is taken to be a spin $-1$, even, ghost number 0, and square zero parameter $\varepsilon^2 = 0. $ We consider Lie algebra cohomology of $\mathfrak{g}[z, \varepsilon]$ relative to the constant Lie subalgebra $\mathfrak{g}$, so that we have
\begin{align}
    \text{Obs}_{\text{HT}}(\mathfrak{g}) = H^*(\mathfrak{g}[z, \varepsilon], \mathfrak{g} ; \,\mathbb{C}).
\end{align}
In the above we identify $\varepsilon = \text{d}z$.

A few remarks before moving on are as follows.

\begin{remark} The identification of local observables as a version of relative Lie algebra cohomology is nicely compatible with papers studying $Q$-cohomology of local observables for the pure holomorphic twist of $\mathcal{N}=4$ Yang-Mills \cite{Chang:2013fba, Chang:2022mjp}, and pure holomorphic twist of $\mathcal{N}=1$ Yang-Mills \cite{Budzik:2023xbr}.
\end{remark}

\begin{remark}
    It's instructive to compare what we've obtained in terms of the original vector multiplet fields. One identifies 
    \begin{align}
    \del^n b &\leftrightarrow D_z^n \,\ov{\lambda}^+_{\dot{2}} \\ \del^{n+1}c &\leftrightarrow D_z^n \lambda^+_{1}. 
    \end{align}  
    Indeed, it is $\del c $ which gets identified with the gaugino $\lambda^+_{1}$ without a covariant derivative, so that the original vector multiplet fields already satisfy the horizontality condition. The $\mathfrak{g}$-invariance condition on the other hand is the familiar condition that a local operator must be a $\mathfrak{g}$-invariant combination of words built from the gauginos $ \,\ov{\lambda}^+_{\dot{2}}$ and $\lambda^+_{1}$ and their holomorphic covariant derivatives.  
\end{remark}

\begin{remark}
We obtained the dg Poisson vertex algebra $\mathcal{C}_{\text{HT}}(\mathfrak{g}) = \big(\mathcal{C}_{bc}(\mathfrak{g}) \big)^{\mathfrak{g}}_{\text{bas}}$ from the \textit{classical} BV formalism. In other words, it is simply is the tree-level complex of observables. One may worry about perturbative corrections to the differential $Q$ and the $\lambda$-bracket $\{ \cdot \,_{\lambda}\, \cdot \}$ coming from the contribution of loop diagrams in perturbation theory (equivalently, from the requirement that at the quantum level, we must satisfy the quantum master equation). Indeed, such corrections do show up in pure holomorphic twists of four-dimensional $\mathcal{N}=1$ theories \cite{Budzik:2023xbr, Budzik:2025zvu}, and in two-dimensional $\mathcal{N}=(0,2)$ theories \cite{Witten:2005px}. Fortunately, in the present case there is an elegant non-renormalization theorem proven in \cite{Balduf:2024wwp} (see also \cite{Wang:2024tjf}) which applies. The latter states that there are no perturbative corrections to the tree-level algebra of observables in a mixed holomorphic-topological theory, provided there are at least two topological directions. Rather remarkably, all conceivable Feynman diagrams beyond tree level that could correct the dg PVA structure simply vanish (individually!)\footnote{The first instance of this phenomenon was noted in the two-dimensional Poisson sigma model, used in the deformation quantization of Poisson manifolds \cite{Kontsevich:1997vb}. In particular, the vanishing of the Feynman diagrams in this example of two topological directions is discussed in Section 6.6 of \cite{Kontsevich:1997vb}.}. Thus the answer obtained above is valid to all orders in perturbation theory.
\end{remark}

\subsection{Recovering the Schur index from the Basic Complex}

Let us finally connect the discussion here with more familiar quantities in the study of four-dimensional $\mathcal{N}=2$ theories. We will show that the character of $\text{Obs}_{\text{HT}}(\mathfrak{g})$ is precisely the Schur index of the theory with a vector multiplet valued in $\mathfrak{g}$. In order to define the character, we recall the gradings on $\mathcal{C}_{\text{HT}}(\mathfrak{g})$. The space $\mathcal{C}_{\text{HT}}(\mathfrak{g})$ is equipped with a ghost number grading, which we recall is \begin{align}
    \text{gh}(\del^k c) = \text{gh}(\del^k b) = 1,
\end{align} along with a spin grading $s$,
\begin{align}
    s(\del^k b) = k+1, \,\,\,\,\, s(\del^k c) = k, 
\end{align} so that we have 
\begin{align}
    \mathcal{C}_{\text{HT}}(\mathfrak{g}) = \bigoplus_{s,g} \mathcal{C}_{\text{HT}}^{s,g}(\mathfrak{g})
\end{align} The differential raises the ghost number by a unit and preserves spin, so that \be Q: \mathcal{C}_{\text{HT}}^{s,g}(\mathfrak{g}) \to \mathcal{C}_{\text{HT}}^{s,g+1}(\mathfrak{g}) .\ee Moreover the subcomplex at a fixed spin $s$ is a finite-dimensional cochain complex. The cohomology $\text{Obs}_{\text{HT}}(\mathfrak{g})$ is then given simply by 
\begin{align}
\text{Obs}_{\text{HT}}(\mathfrak{g}) = \bigoplus_{s,g} H^*(\mathcal{C}^{s,g}_{\text{HT}}(\mathfrak{g})).      
\end{align}  The spin-graded character of $A(\mathfrak{g})$ is then the Euler characteristic
\begin{align}
    \chi_{\text{Obs}_{\text{HT}}(\mathfrak{g})}(q) = \sum_{s,g} \text{dim} \,H^*(\mathcal{C}^{s,g}_{\text{HT}}(\mathfrak{g})  \big)  (-1)^g q^s. 
\end{align} By the Euler-Poincar\'e principle this is simply given by \begin{align}
    \chi_{\text{Obs}_{\text{HT}}(\mathfrak{g})}(q)= \sum_{s,g} \text{dim} \,\mathcal{C}^{s,g}_{\text{HT}}(\mathfrak{g})  \,(-1)^g  q^s. 
\end{align} By definition $\mathcal{C}_{\text{HT}}(\mathfrak{g})$ is given by the $\mathfrak{g}$-invariant part of the exterior algebra in adjoint-valued variables $\del^k c^a $ with $k \geq 1$ and co-adjoint valued variables $\del^k b_a$ with $k \geq 0$. Without imposing $\mathfrak{g}$-invariance, the character would therefore just be the graded character of infinitely many Grassmann variables:
\begin{align}
    \prod_{n \geq 1} (1-q^n)^{2 \text{dim}(\mathfrak{g})} = (q;q)_{\infty}^{2 \text{dim}(\mathfrak{g})}. 
\end{align} The character of the $\mathfrak{g}$-invariant part can be obtained by further refining the above to a $\mathfrak{g}$-character, and projecting to $\mathfrak{g}$-invariants by carrying out a Molien style integral over the Cartan torus of $G$:
\begin{align} \label{schur}
    \chi_{\text{Obs}_{\text{HT}}(\mathfrak{g})}(q) = \frac{1}{|W|} \oint_{|z_i| = 1} \prod_{i=1}^r \frac{dz_i}{ 2\pi i z_i} \,\Delta(z) \,    (q; q)_{\infty}^{2r} \prod_{\alpha >0} (q z^{\alpha};  q )_{\infty}^{2} (q z^{-\alpha}; q)_{\infty}^2  
\end{align} where $r = \text{rk}(G)$, \begin{align} \Delta(z_i) = \prod_{\alpha > 0} (1-z^{\alpha})(1-z^{-\alpha}) \end{align} is the standard Weyl denominator, $W$ is the Weyl group of $\mathfrak{g}$, and $\alpha$ denotes a positive root of $\mathfrak{g}$. This is precisely the vector multiplet contribution to the Schur index provided one identifies the spin $s$ with the fugacity $q$ and the ghost number with the charge under the Cartan of $SU(2)_R$. 

Finally, it is also straightforward to obtain a refined Euler characteristic by incorporating the $B$-number. Recall the $B$-number assignment of the generating fields 
\begin{align}
    B(\del^k c) = 0, \,\,\,\,\, B(\del^k b) = 1.
\end{align} Since the differential $Q$ preserves the $B$-number, we can split the cochain complex $\big(\mathcal{C}_{\text{HT}}(\mathfrak{g}), Q \big)$ as \begin{align}
    \mathcal{C}_{\text{HT}}(\mathfrak{g}) = \bigoplus_{s,g, B} \mathcal{C}_{\text{HT}}^{s,B,g},\,\,\,\,\,\ Q:\mathcal{C}_{\text{HT}}^{s,B,g} \to \mathcal{C}_{\text{HT}}^{s,B,g+1}.
\end{align} The $B$-refined index introduces an additional fugacity $y$, and is given as
\begin{align}
    \chi_{\text{Obs}_{\text{HT}}(\mathfrak{g})}(q,y)= \sum_{s,g} \text{dim} \,\mathcal{C}^{s,g, B}_{\text{HT}}(\mathfrak{g})  \,(-1)^g  y^B q^s. 
\end{align}
Refining the $q$-Pochhammer symbols capturing the character of the exterior algebra accordingly by introducing the fugacity $y$, the $B$-refined Euler characteristic becomes 
\begin{align} \begin{split}
    \chi_{\text{Obs}_{\text{HT}}(\mathfrak{g})}(q,y) =\,\,\,\,\,\,\,\,\,\,\,\,\,\,\,\,\,\,\,\,\,\,\,\,\,\,\,\,\,\,\,\,\,\,\,\,\,\,\,\,\,\,\,\,\,\,\,\,\,\,\,\,\,\,\,\,\,\,\,\,\,\,\,\,\,\,\,\,\,\,\,\,\,\,\,\,\,\,\,\,\,\,\,\,\,\,\,\,\,\,\,\,\,\,\,\,\,\,\,\,\,\,\,\,\,\,\,\,\,\,\,\,\,\,\,\,\,\,\,\,\,\,\,\,\,\,\,\,\,\,\,\,\,\,\,\,\,\,\,\,\,\,\,\,\,\,\,\,\,\,\,\,\,\,\,\,\,\,\,\,\,\,\,\,\,\,\,\,\,\,\,\,\,\,\,\,\,\,\,\,\,\,\,\,\,\,\,\,\,\,\\ \frac{1}{|W|} \oint_{|z_i| = 1} \prod_{i=1}^r \frac{dz_i}{ 2\pi i z_i} \,\Delta(z) \,    (q; q)_{\infty}^{r} (q y ; q)_{\infty}^{r} \prod_{\alpha >0} (q z^{\alpha};  q )_{\infty}(q y z^{\alpha};  q )_{\infty}(q z^{-\alpha}; q)_{\infty}  (q yz^{-\alpha};  q )_{\infty}. \end{split}
\end{align} 

In the literature on four-dimensional $\mathcal{N}=2$ superconformal theories this refinement is known as the Macdonald index with the $y$ variable conventionally written as $T$. However, as discussed above, the $B$-refined character can be defined without making any assumption about conformal invariance.

\subsection{The Case $\mathfrak{g}= \mathfrak{sl}_2$} Having formulated the space of holomorphic-topological observables $\text{Obs}_{\text{HT}}(\mathfrak{g})$ as the $Q$-cohomology of $\mathfrak{g}$-basic elements of the $bc$-ghost system, we now discuss the primary case of interest to us, namely $\mathfrak{g}= \mathfrak{sl}_2$. 

We recall the Poisson vertex algebra $A$ discussed in detail in Section 2. $A$ is by definition a quotient Poisson vertex algebra $V/I$, where $V$ has generators consisting of a commuting spin 2, ghost number 2 field $X$, and an anti-commuting ghost number $3$ and spin $3$ field $Y$, \be V = \mathbb{C}[X, \del X, \del^2 X, \dots ] \otimes \Lambda[ Y, \del Y, \del^2 Y, \dots],\ee equipped with the $\lambda$-brackets 
\begin{align}
    \{X \,_{\lambda} \,X\} &= \del X + 2 \lambda X, \\ \{X \,_{\lambda} \, Y\} &= \del Y + 3 \lambda Y, \\ \{Y\,_{\lambda}\, Y\} &= 0. 
\end{align} The Poisson vertex ideal $I$ is defined to be the smallest Poisson vertex ideal containing the element $X^2$.

We now construct a homomorphism of Poisson vertex algebras 
\begin{align}
    \varphi: A \to \text{Obs}_{\text{HT}}(\mathfrak{sl}_2), 
\end{align} as follows. We identify $\mathfrak{sl}_2$ with $\mathfrak{so}_3$, so that the elements of $\mathcal{C}_{\text{HT}}(\mathfrak{so}_3)$ are given $\mathfrak{so}_3$-invariant products of $b^i, \del b^i, \dots$ (after identifying the dual of $\mathfrak{so}_3$ with $\mathfrak{so}_3$ via the invariant pairing) and $\del c^i, \del^2 c^i$ for $i=1,2,3$. The $\mathfrak{so}_3$-invariant words are then built from the tensors $\delta_{ij}$ and $\epsilon_{ijk}.$ We set  
\begin{align}
    \varphi(X) &= -[\delta_{ij}\, b^i \del c^j], \\
    \varphi(Y) &= [\epsilon_{ijk} \, b^i b^j b^k],
\end{align} extended on all of $A$ by requiring $[\varphi, \del ] =0$ and by requiring \begin{align}
    \varphi(a b) = \varphi(a) \varphi(b).
\end{align}

For the map $\varphi$ to be sensible, we must first check that that $\delta_{ij} b^i \del c^j$ and $\epsilon_{ijk} b^i b^j b^k$ are $Q$-closed elements of $\mathcal{C}_{\text{HT}}(\mathfrak{sl}_2)$. That this is the case can be seen without any calculations in fact: $Q$ must map a basic element to a basic element, and so $Q(\delta_{ij} \, b^i \del c^j)$ must be a basic element of spin $2$, built from one $b$ and two $c$'s. But there are no such elements (the minimal spin element of this type would be $\epsilon_{ijk} b^i \del c^j \del c^k$ however, this has spin $3$). Similarly, $Q(\epsilon_{ijk} b^i b^j b^k)$ must be a spin $3$ basic element built from 3 $b$'s and a single $c$ and again, there are no such elements (the minimal spin element of this type is $\delta_{ij} \delta_{kl} b^i b^j b^k \del c^l $ but this has spin $4$). Thus 
\begin{align}
Q(\delta_{ij} \, b^i \del c^j) &= 0, \\
Q(\epsilon_{ijk} \, b^i b^j b^k) &= 0,
\end{align} and the map is well-defined. Since $\del$ commutes with $Q$, this also shows that all derivatives are $Q$-closed elements, 
\begin{align}
    Q \big(\del^k (\delta_{ij} \, b^i \del c^j) \big) &= 0, \\  Q \big(\del^k(\epsilon_{ijk} \, b^i b^j b^k) \big) &= 0. 
\end{align} 

To check whether the super-commutative product is preserved under $\varphi$, observe that $\varphi(X) = -[\delta_{ij} b^i \del c^j]$ indeed behaves as an even commutative element and whereas $\varphi(Y) = [\epsilon_{ijk} b^i b^j b^k]$ behaves as an odd anti-commuting element. 

Next we must check that the $\lambda$-bracket is preserved, which amounts to checking the equations
\begin{align}
\varphi(\{ X \,_{\lambda} X \}) &= \{ \varphi(X) \,_{\lambda} \, \varphi(X) \} , \\
\varphi(\{ X \,_{\lambda} Y \}) &= \{ \varphi(X) \,_{\lambda} \, \varphi(Y) \}, \\
\varphi(\{ Y \,_{\lambda} Y \}) &= \{ \varphi(Y) \,_{\lambda} \, \varphi(Y) \}.
\end{align}
The $XX$ equation goes as follows. We must compute 
\begin{align}
    \{\delta_{ij} b^i \del c^j \,_{\lambda} \, \delta_{kl} b^k \del c^l \}. 
\end{align} A straightforward calculation applying the left and right Leibniz rules \eqref{leib1}, \eqref{leib2} shows that 
\begin{align}
\{[\delta_{ij} b^i \del c^j]\,_{\lambda} \, [\delta_{kl} b^k \del c^l] \} &= (\del ([-\delta_{ij} b^i \del c^j] ) + 2 \lambda [ -\delta_{ij} b^i \del c^j] ) , \\ &= \del (\varphi(X)) + 2\lambda (\varphi(X)) , \\ &= \varphi( \del X + 2\lambda X) = \varphi(\{X \, _{\lambda} \,X \}). 
\end{align} Similarly to check the preservation of the $XY$ bracket, one can calculate 
\begin{align}
    \{ [-\delta_{ij} b^i \del c^j] \,_{\lambda} \, [\epsilon_{klm} b^k b^l b^m] \} &= \del([\epsilon_{klm} b^k b^l b^m]) + 3\lambda [\epsilon_{klm} b^k b^l b^m], \\
    &= \del \varphi(Y) + 3 \lambda \varphi(Y)  \\ &= \varphi( \del Y + 3\lambda Y) = \varphi(\{X \,_{\lambda} Y \}).
\end{align} Finally the $YY$ bracket is preserved because the left hand side is simply zero, as is the case for the right hand-side since the $\lambda$-bracket of $b$ vanishes with itself. 

Finally, in order to check that $\varphi$ is indeed a homomorphism from $A$ (and not $V$), we must check that 
\begin{align}
    \varphi(X^2) = 0. 
\end{align} We have 
\begin{align}
    \varphi(X^2) = [\delta_{ij} b^i \del c^j \, \delta_{kl} b^k \del c^l]. 
\end{align} This vanishes if one can show that the element $\delta_{ij} b^i \del c^j \, \delta_{kl} b^k \del c^l$ on the right-hand-side is $Q$-exact. This indeed holds, since one can check
\begin{align}
    Q(-\frac{1}{2} \epsilon_{ijk} b^i b^j  \,\del^2 c^k) = \delta_{ij} b^i \del c^j \delta_{kl} b^k \del c^l.
\end{align} Therefore one has 
\begin{align}
    \varphi(X^2) = 0,
\end{align} and the homomorphism $\varphi$ is established. 

In fact, we expect that $\varphi$ is an \textit{isomorphism} of Poisson vertex algebras.

\begin{conjecture}
    The map $ \varphi: A \to \text{Obs}_{\text{HT}}(\mathfrak{sl}_2)$ is an isomorphism of Poisson vertex algebras.
\end{conjecture}

We wish to establish the claim that $\varphi$ is an isomorphism in future work. In the present note we simply perform some non-trivial checks of this claim. The first non-trivial check is to verify that the spin-graded Euler characteristics of $A$ and $\text{Obs}_{\text{HT}}(\mathfrak{g})$ agree. The Euler characteristic of $A$, by definition is the specialization of the Hilbert-Poincar\'e series $P_A(t,q)$ at $t=-1$. Recalling from Section $2$ that
\begin{align}
    P_A(q,t) = \sum_{n =0}^{\infty} \frac{(-tq ;q)_n}{(q; q)_n} t^{2n} q^{n(n+1)},
\end{align} the specialization at $t=-1$ gives
\begin{align}
    \chi_A(q) = \sum_{n =0}^{\infty} q^{n(n+1)}.
\end{align}
On the other hand, the Euler characteristic of $\text{Obs}_{\text{HT}}(\mathfrak{sl}_2)$ was established to be given by the integral \eqref{schur}. Specializing to $\mathfrak{g}= \mathfrak{sl}_2$ gives us the integral
\begin{align}
    \chi_{\text{Obs}_{\text{HT}}(\mathfrak{sl_2})}(q) = \frac{1}{2} \oint_{|z| = 1} \frac{dz}{(2\pi i z)} (1-z^2)(1-z^{-2})(q;q)^2_{\infty}(q z^2; q)^2_{\infty} (q z^{-2}; q)^2_{\infty}.
\end{align}The Jacobi triple product identity allows us to rewrite
\begin{equation}
    (q;q)_{\infty} (q z^2 ; q)_{\infty}(q z^{-2}; q)_{\infty} = \frac{1}{1-z^{-2}} \sum_{n \in \mathbb{Z}} (-1)^n z^{2n} q^{\frac{n(n+1)}{2}},
\end{equation}
which immediately yields
\begin{align}
    \chi_{\text{Obs}_{\text{HT}}(\mathfrak{sl_2})}(q) = \sum_{n =0}^{\infty} q^{n(n+1)},
\end{align} thus establishing the equality of spin-graded Euler characteristics 
\begin{align}
    \chi_A(q) =  \chi_{\text{Obs}_{\text{HT}}(\mathfrak{sl_2})}(q).
\end{align}
As an even more non-trivial check, one can try to compare the $y$-refined Euler character. Specializing the $y$-refined Hilbert-Poincar\'e series of $A$ which we recall was established to be
\begin{align}
    P_A(t,q,y) = \sum_{n = 0}^{\infty} q^{n(n+1)} t^{2n}y^{n} \frac{(-t q y^2 ; q)_n}{(q; q)_n},
\end{align} to $t=-1$ gives
\begin{align} \label{yrefa}
    \chi_A(q,y) = \sum_{n=0}^{\infty} q^{n(n+1)} y^{n} \frac{(  y^2 q ; q)_n}{(q; q)_n}.
\end{align}
On the other hand, the $y$-refined character of $\text{Obs}_{\text{HT}}(\mathfrak{sl}_2)$ is given by the contour integral 
\begin{multline}
\chi_{\mathrm{Obs}_{\mathrm{HT}}(\mathfrak{sl}_2)}(q,y) =\\
 -\frac{1}{2} \oint_{|z|=1} \frac{dz}{2\pi i z}\,
(z-z^{-1})^2
(q;q)_{\infty}(q z^2;q)_{\infty}(q z^{-2};q)_{\infty}
(q y;q)_{\infty}(q y z^2;q)_{\infty}(q y z^{-2};q)_{\infty}.
\end{multline} Though obtaining a closed-form for the constant term extraction enforced by the contour integral seems non-trivial\footnote{On the other hand, the conjectured isomorphism of $\varphi$ would imply the nice ``constant term" identity \begin{align}-\frac{1}{2} \oint_{|z|=1} \frac{dz}{2\pi i z}\,
(z-z^{-1})^2
(q;q)_{\infty}(q z^2;q)_{\infty}(q z^{-2};q)_{\infty}
(q y;q)_{\infty}(q y z^2;q)_{\infty}(q y z^{-2};q)_{\infty} = \sum_{n \geq 0} q^{n(n+1)} y^{n} \frac{(y^2 q ; q)_n}{(q; q)_n} \end{align} as a Corollary. }, it is straightforward to obtain a term by term series of the form $\sum_{n \geq 0} P_n(y) q^n$ to a given spin $S$. One can perform this extraction and check the agreement with \eqref{yrefa} to high spin (e.g we checked for $S=500$), giving us 
\begin{align}
    \chi_A(q,y) - \chi_{\text{Obs}_{\text{HT}}(\mathfrak{sl_2})}(q,y) = O(q^{S}), \,\,\,\, S \gg 1. 
\end{align}

Of course, it's most conclusive to compare the Hilbert-Poincar\'e series of $A$ and $\text{Obs}_{\text{HT}}(\mathfrak{sl}_2)$ directly, with no specializations. Since the latter, as a graded vector space is defined as a relative Lie algebra cohomology, which at a fixed spin is defined as the cohomology of a finite-dimensional cochain complex, one can calculate the Hilbert-Poincar\'e series mod a finite, computationally manageable spin $S$ with the aid of computational linear algebra. One simply implements the basic $bc$-complex equipped with the Chevalley-Eilenberg differential and computes its cohomology by using computer linear algebra techniques. We performed the check for spins $S \leq 20$ finding agreement to this order\footnote{With enough optimization, this verification takes a few hours of runtime on a standard PC at the time of writing.}:
\begin{align}
    P_A(t,q,y) - P_{\text{Obs}_{\text{HT}}(\mathfrak{sl_2})}(t,q,y) = O(q^{21}).
\end{align}
\section{The Differential $Q_{\text{inst}}$ and its Cohomology} \label{np} Having established the Poisson vertex algebra $A$ as a candidate for the space of holomorphic-topological observables $\text{Obs}_{\text{HT}}(\mathfrak{sl}_2)$ of the pure $\mathcal{N}=2$, $SU(2)$ gauge theory to all orders in perturbation theory, it is natural to wonder whether non-perturbative effects can modify the result. In this section we observe that $A$ admits a natural differential \begin{align}
    Q_{\text{inst}}: A^{s, \text{gh}} \to A^{s, \text{gh}+1}, \,\,\,\, (Q_{\text{inst}})^2 =0,
\end{align} preserving the spin $s$ and raising the ghost number by $+1$, and compute the cohomology of this differential. We will see that $Q_{\text{inst}}$ can radically modify the result, lifting most of the observables and leaving only a single, one-dimensional space at the special spins 
\begin{align}
    s_n = n(n+1), \,\,\,\,\, n=0,1,2, \dots.
\end{align}
As the notation suggests, we hypothesize that $Q_{\text{inst}}$ arises from instanton effects in the underlying $\mathcal{N}=2$ gauge theory. 

Though we do not delve into this point much further in the present article, there's a simple reason to expect a non-trivial contribution to the differential from non-perturbative effects. The perturbative differential $Q_{\text{HT}}$ acts on the $b$-ghost via 
\begin{align}
Q b_a = f^a_{bc} \, c^b b^c,    
\end{align} and thus manifestly preserves the $B$-number. Thus the $Q$-cohomology is graded by the spin $s$, the ghost number $g$, and this additional $B$-number. In terms of the symmetry generators of the $\mathcal{N}=2$ algebra, the first two of these we recall are
\begin{align}
    s &= J_{12} + I_3, \\
    g &= I_3,
\end{align} where $I_3$ is the Cartan of the $SU(2)_R$ symmetry, and $J_{12}$ the rotation generator in the $(12)$-plane. The $B$-number symmetry on the other hand can be traced back to the fact that the classical theory possesses a $U(1)_r$ symmetry. More precisely, the identification between the $U(1)_r$-charge and the quantum numbers of the holomorphic-topological twist is 
\begin{align}
    B = \frac{1}{2}(I_3 - r),
\end{align} equivalently,
\begin{align}
    r = g-2B.
\end{align}  For instance with this identification, one indeed has \begin{align}
    r(\del c) = 1, \,\,\,\, r(b) = -1,
\end{align} corresponding to the $r$-charges of the two gauginos $\lambda^+_{1}$ and $\ov{\lambda}^+_{\dot{2}}.$ 
However, this $U(1)_r$ is broken by instanton effects, and thus at a non-perturbative level there's no reason to expect that the differential $Q$ preserves the $B$-number. 

We therefore posit a differential $Q_{\text{inst}}: V \to V$ of the form 
\begin{align}
Q_{\text{inst}}(X) &= 0, \\
    Q_{\text{inst}}(\del^k X) &= k \,\del^{k-1}Y,  \,\,\,\,\,\, k \geq 1, \\ 
    Q_{\text{inst}}(\del^k Y) &= 0 \,\,\,\,\,\,\,\,\,\,\,\,\,\,\,\,\,\,\,\,\,\,\,\,\, k \geq 0,
\end{align} and extend it to act on all of $V$ by requiring $Q_{\text{inst}}$ to be an odd derivation of the super-commutative product on $V$. Defined as such $Q_{\text{inst}}$ still preserves the spin $s$ and the ghost number but now carries a $B$-charge of $+2$, \begin{align}
    B(Q_{\text{inst}}) = 2.
\end{align}Observe also that defined as such $Q_{\text{inst}}$ preserves the differential ideal $I$. 

To see the preservation of the ideal $I$ it is convenient to go back to the generating function form of the fields. Letting $X(z), Y(z)$ be as in Section \ref{PVA}, recall that the ideal $I$ is generated by coefficients of 
\begin{align}
    X(z)^2, \,\,\, X(z) Y(z), \,\,\,, X(z) \del Y(z), \,\,\, Y(z) \del Y(z).
\end{align} In terms of generating functions we have 
\begin{align}
    Q_{\text{inst}} \big(X(z) \big) = z Y(z), 
\end{align} so that
\begin{align}
    Q_{\text{inst}}\big(X(z)^2 \big)  &= 2z X(z) Y(z), \\ 
    Q_{\text{inst}}\big(X(z) Y(z) \big) &= 0, \\ 
    Q_{\text{inst}} \big(X(z) \del Y(z) \big) &= z \,Y(z) \del Y(z), \\
    Q_{\text{inst}} \big(Y(z) \del Y(z) \big) &= 0,
\end{align} so that the ideal $I$ is indeed preserved by $Q_{\text{inst}}$. Thus $Q_{\text{inst}}$ descends to a well-defined differential on the quotient \begin{align} Q_{\text{inst}}:A \to A.\end{align} Observe however, that $Q_{\text{inst}}$ does not commute with the derivation $\del$ on $A$, nor is it a derivation of the $\lambda$-bracket. Thus the $Q_{\text{inst}}$-cohomology is only guaranteed to be a supercommutative algebra. 

We now compute the cohomology $H^*_{Q_{\text{inst}}}(A)$. The claim is as follows. The cohomology is only non-vanishing at even ghost numbers $\text{gh}=2n$, where it is one-dimensional, spanned by a cohomology class of spin $s_n = n(n+1)$. The non-trivial cohomology class at $\text{gh}=2n$ is 
\begin{align}
    \alpha_{2n} := [X \,\del^2X \, \del^4 X \dots \del^{2n-2}X].
\end{align} In order to prove this, we go back to the superfield description of Section \ref{PVA}. Recalling that \begin{align}
    \mathcal{K}(z, \theta) = X(z) + \theta \,Y(z),
\end{align} the differential $Q_{\text{inst}}$ takes the simple form \begin{align}
    Q_{\text{inst}} \big(\mathcal{K}(z,\theta) \big) = z \frac{\del}{\del \theta}\big(\mathcal{K}(z,\theta) \big).
\end{align} In order to compute the cohomology then, it is useful to work in the graded dual space $A^{\vee}$. The latter was established as taking the following form 
\begin{align}
    A^{\vee} = \bigoplus_{n=0}^{\infty} \Delta_n^2 (\Omega(\mathbb{C}^n))^{S_n},
\end{align} namely a direct sum over all $n$ of the space of invariant polynomial differential forms in $z_1, \dots, z_n$ multiplied by an overall factor given by the Vandermonde squared \be (\Delta(z_1, \dots, z_n))^2  = \prod_{i < j}(z_i -z_j)^2.\ee Let us consider a length $n$ dual, namely a polynomial differential form $\eta$ obtained via \begin{align}
    \omega(z_1, \dots, z_n, \theta_1, \dots, \theta_n) = \eta \big(\mathcal{K}(z_1, \theta_1) \dots \mathcal{K}(z_n, \theta_n) \big)
\end{align} The dual differential $Q_{\text{inst}}^{\vee}$, defined by \be Q_{\text{inst}}^{\vee} \eta = \eta Q_{\text{inst}} \ee therefore simply takes an element $\omega \in  \Delta_n^2 (\Omega(\mathbb{C}^n))^{S_n}$ and acts on it via $\sum_{i=1}^n z_i \frac{\del}{\del \theta_i}$, 
\begin{align}
    Q^{\vee}_{\text{inst}}(\omega) = \sum_{i=1}^n z_i \frac{\del}{\del \theta_i} \omega
\end{align} which in the language of differential forms simply is the contraction operator $\iota_V $ for the vector field $V = \sum_{i=1}^n z_i \frac{\del}{\del z_i},$ 
\begin{align}
    Q_{\text{inst}}^{\vee} (\omega) = \iota_V \omega,\,\,\,\,\,\,\,\,\, V = \sum_{i=1}^n z_i \frac{\del}{\del z_i}.
\end{align} Letting $e_1, \dots, e_n$ be the elementary symmetric polynomials in $(z_1, \dots, z_n)$ as in Section \ref{PVA}, recall that the invariant differential forms are given as 
\begin{align}
    \Omega(\mathbb{C}^n)^{S_n} \cong \mathbb{C}[e_1, \dots, e_n] \otimes \Lambda[\chi_1, \dots, \chi_n]
\end{align} where \begin{align}
    \chi_i = de_i.
\end{align} Now note that on $f$ a homogeneous polynomial of degree $k$, we have \be \mathcal{L}_V f = \sum_{i=1}^n z_i \frac{\del}{\del z_i} f = k f,\ee therefore 
\begin{align}
    \iota_V (df) = \mathcal{L}_V f  = k f. 
\end{align} In particular, since $e_k$ is homogeneous of degree $k$ we have
\begin{align}
    Q_{\text{inst}}^{\vee} (de_k) = k e_k
\end{align}
Thus on an $S_n$-invariant differential form $\mathbb{C}[e_1, \dots, e_n] \otimes \Lambda[\chi_1, \dots, \chi_n]$ the dual differential $Q_{\text{inst}}^{\vee}$ simply acts by
\begin{align}
    Q_{\text{inst}}^{\vee} = \sum_{k=1}^n k \,e_k \frac{\del}{\del \chi_k}.
\end{align} The cochain complex 
\begin{align}
     \Big(\mathbb{C}[e_1, \dots, e_n] \otimes \Lambda[\chi_1, \dots, \chi_n] \, , \,  Q^{\vee}_{\text{inst}} = \sum_{k=1}^n k \,e_k \frac{\del}{\del \chi_k} \Big)
\end{align}
we've obtained is nothing but the Koszul complex for the regular sequence 
\begin{align}(e_1, 2e_2, \dots, ne_ n)\end{align} in the polynomial ring $\mathbb{C}[e_1, \dots, e_n]$, and therefore simply has cohomology concentrated in degree $0$, generated by the class of constant zero-form $1$. Since $Q_{\text{inst}}^{\vee}$ simply commutes through the Vandermonde square factor, we obtain that the only surviving cohomology class in $A_n^{\vee}$ is given by the invariant zero-form 
\begin{align}
    \eta_{2n} = [ \Delta(z_1, \dots,z_n)^2 ] .
\end{align} Since $z_i$ has spin $-1$, the Vandermonde factor has spin $-n(n-1)$ and taking the base spin and ghost numbers of $-2n$ into account, we thus see that the dual complex $(A_n^{\vee}, Q^{\vee}_{\text{inst}})$ has a one-dimensional cohomology at each ghost number $-2n$ and each spin $-n(n+1)$. Dualizing back to $A$, we find that the cohomology is one-dimensional at ghost number $2n$ and consists of a single generator of spin $n(n+1)$. The unique element in $A$ with these degrees is given by 
\begin{align}
    \alpha_{2n} = [X \,\del^2X \dots \del^{2n-2}X],
\end{align} thus establishing the claim.

The Hilbert-Poincar\'e series for $H^*_{Q_{\text{inst}}}(A)$ is thus
\begin{align} \label{hpinst}
    P_{H^*_{Q_{\text{inst}}}(A)}(t,q) = \sum_{n=0}^{\infty} t^{2n} q^{n(n+1)}.
\end{align}

To close this section, it is useful to translate $Q_{\text{inst}}$ back to the vector multiplet fields. Observing that under the morphism $\varphi: A \to \text{Obs}_{\text{HT}}(\mathfrak{sl}_2)$ (assuming invertibility) the differential $Q_{\text{inst}}$ becomes
\begin{align}
    \hat{Q}_{\text{inst}}\big([ \delta_{ij} b^i(z) \del_z c^j(z)] \big) = -[ \epsilon_{ijk} \,z\, b^i(z) b^j(z) b^k(z) ].
\end{align}
Recalling that $b$ and $\del c$ are identified with the gauginos $\lambda^+_{1}$ and $\ov{\lambda}^+_{\dot{2}}$ of the $\mathcal{N}=2$ vector multiplet, this suggests that the full form of the holomorphic-topological differential in the $\mathcal{N}=2$ theory takes the form
\begin{align} \label{qht}
    Q_{\text{HT}} = Q_{\text{pert}} + \Lambda^2 \, Q_{\text{inst}} 
\end{align} where $Q_{\text{pert}}$ is the standard BRST differential as discussed in Section \ref{sw},  $\Lambda$ denotes the dynamical scale of the pure $\mathcal{N}=2$, $SU(2)$ theory, and
\begin{align}
    Q_{\text{inst}}\big(\text{Tr}(\ov{\lambda}(z) \lambda(z)) \big) = a z\,\text{Tr}\big([ \,\ov{\lambda}(z), \ov{\lambda}(z)]\ov{\lambda}(z) \big).
\end{align} Here $a$ is an $r$-neutral constant. The reason the particular factor $\Lambda^2$ is introduced is that if we remember that $\Lambda$ carries $r$-charge $+2$, and $Q_{\text{HT}}$ carries $r$-charge $1$,
\begin{align}
    r(\Lambda) = 2, \,\,\,\,\, r(Q_{\text{HT}}) = 1,
\end{align}
then the $r$-charges of both sides of the equation \ref{qht} agree.  It would be very interesting to understand better the origins of such a formula from non-perturbative physics. 

\section{Future Directions} \label{future} 
We believe the following points are worth addressing in future work.

\begin{itemize}
    \item While we have provided ample evidence for the claimed isomorphism 
    \begin{align} \begin{split}
        H^*\big(\mathfrak{sl}_2[[z]], \mathfrak{sl}_2 ; \, \Lambda\big((\mathfrak{sl}_2[[z]] \,\text{d}z)^{\vee}\big) \big) \cong \,\,\,\,\,\,\,\,\,\,\,\,\,\,\,\,\,\,\,\,\,\,\,\,\,\,\,\,\,\,\,\,\,\,\,\,\,\,\,\,\,\,\,\,\,\,\,\,\,\,\,\,\,\,\,\,\,\,\,\,\,\,\,\,\,\,\,\,\,\,\,\,\,\,\,\,\,\,\,\,\, \\ \mathbb{C}[X, \del X, \dots]\otimes \Lambda[Y, \del Y, \dots]/\langle X^2,\; XY,\; X\partial Y,\; Y\partial Y\rangle_{\partial},\end{split}
    \end{align}
    proving this rigorously is an important open problem of independent mathematical interest. It would be interesting to see whether the techniques of \cite{fgt} can be adopted to prove this claim. \\
    \item In Section \ref{np} we posited the differential $Q_{\text{inst}}$ mainly from algebraic considerations. It would be quite interesting to understand the origins of this from a one-instanton calculation in the pure $\mathcal{N}=2$, $SU(2)$ gauge theory. In particular we find it a curious state of affairs that our proposed differential $Q_{\text{inst}}$, though mathematically quite natural, does not commute with holomorphic translations, but instead satisfies a formula for the form \begin{align} \big[ \del_z , Q_{\text{inst}} \big] = \frac{\del}{\del \theta} .\end{align} \\ 
    \item Given the fact that the algebra of bulk holomorphic-topological observables for the special case of $\mathfrak{g}= \mathfrak{sl}_2$ seems to admit such an explicit characterization, it's natural to wonder to what extent this holds for more general Lagrangian $\mathcal{N}=2$ gauge theories. It would also be quite interesting to incorporate line defects, surface defects and boundary conditions and to see whether the resulting spaces are also explicitly computable.  \\

 \item Finally, it would be very illuminating to have a detailed comparison of the algebras of observables discussed in this note to calculations performed in the infrared effective theory. Can one reproduce either the Poisson vertex algebra $A$, or the more speculative $H^*_{Q_{\text{inst}}}(A)$ (or both somehow) from a calculation involving only the abelian vector multiplets and the spectrum of BPS particles at a given point on the Coulomb branch? \\

If one formally applies the rules of \cite{Andrews:2025tko}, which in the superconformal case gives a prescription to recover the Macdonald index $\chi(q,y)$ from a BPS quiver, to the 2-Kronecker quiver (which captures the BPS spectrum of the Seiberg-Witten $SU(2)$ theory in the strong coupling chamber), one obtains
\begin{align}
(q;q)_{\infty}(qy; q)_{\infty} \text{Tr} (\mathcal{O}(q,y)) = 1+y q^2 + y^2 q^6 + y^3 q^{12} + \dots.
\end{align}
This is decidedly different from the perturbative Macdonald index i.e the $B$-refined character $\chi_A(q,y)$ \eqref{yrefa} of $A$! However, rather encouragingly, the result precisely agrees with the character of $H^*_{Q_{\text{inst}}}(A)$ (once we restore $y$) \eqref{hpinst}. This seems to suggest to us that additional differentials such as $Q_{\text{inst}}$, however they may arise, are crucial if one is to match calculations performed in the ultraviolet to those in the infrared. 
\end{itemize}

\appendix

\section{AI Assisted Exploration} \label{gpt}

In this appendix we briefly record one instance in which a frontier language model
(\texttt{GPT 5.2-Pro}) proved useful during the exploratory phase of this project.
The goal was not to obtain proof, but to accelerate the cycle of conjecture generation
and falsification for the relative Lie algebra cohomology
\be
H^\ast\!\left(\mathfrak{sl}_2[[z]], \mathfrak{sl}_2 \, ; \, \Lambda\bigl((\mathfrak{sl}_2[[z]]\,\text{d}z)^\vee\bigr)\right).
\ee
Since the cochain complex at fixed spin is finite-dimensional, this problem is
particularly well-suited to finite-spin experimentation: one may compute low-spin
cohomology by explicit linear algebra and compare the resulting data against candidate
structural descriptions.

The workflow we used was as follows:
\begin{enumerate}
    \item Formulate the finite-spin cohomology problem in a way suitable for symbolic
    or computational experimentation.
    \item Ask the model to propose a candidate description of the cohomology in terms of say, generators and relations.
    \item Extract concrete finite-spin predictions from that proposal.
    \item Compare those predictions against an independent brute-force cohomology computation.
    \item Refine or discard the proposal accordingly.
\end{enumerate}

In the present case, the model first proposed an incorrect description of the
cohomology in terms of two towers of generators and a preliminary set of relations.
That proposal was then tested against a brute-force computation of the
Hilbert--Poincar\'e series through spin $12$, where it failed. After being confronted
with the mismatch, the model suggested a refined family of relations, which were then
checked independently in the next few spins and served as the starting point for the
mathematical developments pursued in the main text.

The point of recording this example is not that the model supplied proof. Rather, its
value was in accelerating the generation of candidate patterns and in making it easier
to iterate between the finite-spin computation and closed-form conjectures for the full cohomology. In our view, this is a natural and useful mode of AI-assisted exploration for problems in which low-degree
or low-spin data can be computed independently and used as a stringent check on closed-form
conjectural descriptions.

The full transcript of the interaction is available upon request.


\begin{thebibliography}{99}

\bibitem[AKSZ97]{Alexandrov:1995kv}
M.~Alexandrov, A.~Schwarz, O.~Zaboronsky and M.~Kontsevich,
``The Geometry of the master equation and topological quantum field theory,''
Int. J. Mod. Phys. A \textbf{12}, 1405-1429 (1997)
doi:10.1142/S0217751X97001031
[arXiv:hep-th/9502010 [hep-th]].

\bibitem[ABKT25]{Andrews:2025tko}
G.~Andrews, A.~Banerjee, R.~K.~Singh and R.~Tao,
``Macdonald Index From Refined Kontsevich-Soibelman Operator,''
[arXiv:2511.07521 [hep-th]].

\bibitem[BG25]{Balduf:2024wwp}
P.~H.~Balduf and D.~Gaiotto,
``Combinatorial proof of a non-renormalization theorem,''
JHEP \textbf{05}, 120 (2025)
doi:10.1007/JHEP05(2025)120
[arXiv:2408.03192 [math-ph]].

\bibitem[BBBDN20]{Beem:2018fng}
C.~Beem, D.~Ben-Zvi, M.~Bullimore, T.~Dimofte and A.~Neitzke,
``Secondary products in supersymmetric field theory,''
Annales Henri Poincare \textbf{21}, no.4, 1235-1310 (2020)
doi:10.1007/s00023-020-00888-3
[arXiv:1809.00009 [hep-th]].

\bibitem[BLLWRR13]{Beem:2013sza}
C.~Beem, M.~Lemos, P.~Liendo, W.~Peelaers, L.~Rastelli and B.~C.~van Rees,
``Infinite Chiral Symmetry in Four Dimensions,''
Commun. Math. Phys. \textbf{336}, no.3, 1359-1433 (2015)
doi:10.1007/s00220-014-2272-x
[arXiv:1312.5344 [hep-th]].

\bibitem[BGKWWY24]{Budzik:2023xbr}
K.~Budzik, D.~Gaiotto, J.~Kulp, B.~R.~Williams, J.~Wu and M.~Yu,
``Semi-chiral operators in 4d $ \mathcal{N} $ = 1 gauge theories,''
JHEP \textbf{05}, 245 (2024)
doi:10.1007/JHEP05(2024)245
[arXiv:2306.01039 [hep-th]].

\bibitem[BK25]{Budzik:2025zvu}
K.~Budzik and J.~Kulp,
``Loop Corrected Supercharges from Holomorphic Anomalies,''
[arXiv:2512.07771 [hep-th]].

\bibitem[B20-I]{Butson:2020coe}
D.~Butson,
``Equivariant localization in factorization homology and applications in mathematical physics I: Foundations,''
[arXiv:2011.14988 [math.RT]].

\bibitem[B20-II]{Butson:2020mmu}
D.~Butson,
``Equivariant localization in factorization homology and applications in mathematical physics II: Gauge theory applications,''
[arXiv:2011.14978 [math.RT]].

\bibitem[CW19]{cw}
S.~Cautis and H.~Williams,
\newblock ``Cluster theory of the coherent Satake category,''
\newblock { Journal of the American Mathematical Society}, 32(3):709--778,
  2019. 
  [arXiv:1801.08111v2 [math.RT]].
  
\bibitem[CNV10]{Cecotti:2010fi}
S.~Cecotti, A.~Neitzke and C.~Vafa,
``R-Twisting and 4d/2d Correspondences,''
[arXiv:1006.3435 [hep-th]].

\bibitem[CSVY17]{Cecotti:2015lab}
S.~Cecotti, J.~Song, C.~Vafa and W.~Yan,
``Superconformal Index, BPS Monodromy and Chiral Algebras,''
JHEP \textbf{11}, 013 (2017)
doi:10.1007/JHEP11(2017)013
[arXiv:1511.01516 [hep-th]].

\bibitem[CY13]{Chang:2013fba}
C.~M.~Chang and X.~Yin,
``1/16 BPS states in $\mathcal N=$ 4 super-Yang-Mills theory,''
Phys. Rev. D \textbf{88}, no.10, 106005 (2013)
doi:10.1103/PhysRevD.88.106005
[arXiv:1305.6314 [hep-th]].

\bibitem[CL23]{Chang:2022mjp}
C.~M.~Chang and Y.~H.~Lin,
``Words to describe a black hole,''
JHEP \textbf{02}, 109 (2023)
doi:10.1007/JHEP02(2023)109
[arXiv:2209.06728 [hep-th]].

\bibitem[CGS16]{Cordova:2016uwk}
C.~Cordova, D.~Gaiotto and S.~H.~Shao,
``Infrared Computations of Defect Schur Indices,''
JHEP \textbf{11}, 106 (2016)
doi:10.1007/JHEP11(2016)106
[arXiv:1606.08429 [hep-th]].

\bibitem[CS16]{Cordova:2015nma}
C.~Cordova and S.~H.~Shao,
``Schur Indices, BPS Particles, and Argyres-Douglas Theories,''
JHEP \textbf{01}, 040 (2016)
doi:10.1007/JHEP01(2016)040
[arXiv:1506.00265 [hep-th]].

\bibitem[Cos13]{Costello:2013zra}
K.~Costello,
``Supersymmetric gauge theory and the Yangian,''
[arXiv:1303.2632 [hep-th]].

\bibitem[Ded23]{Dedushenko:2023cvd}
M.~Dedushenko,
``On the 4d/3d/2d view of the SCFT/VOA correspondence,''
[arXiv:2312.17747 [hep-th]].

\bibitem[DWP25]{Dimofte:2025oqf}
T.~Dimofte, W.~Niu and V.~Py,
``Line Operators in 3d Holomorphic QFT: Meromorphic Tensor Categories and dg-Shifted Yangians,''
[arXiv:2508.11749 [hep-th]].

\bibitem[ESW22]{Elliott:2020ecf}
C.~Elliott, P.~Safronov and B.~R.~Williams,
``A taxonomy of twists of supersymmetric Yang{\textendash}Mills theory,''
Selecta Math. \textbf{28}, no.4, 73 (2022)
doi:10.1007/s00029-022-00786-y
[arXiv:2002.10517 [math-ph]].


\bibitem[FGT08]{fgt}  
     S.~Fishel, I.~Grojnowski and C.~Teleman, 
     ``The strong {M}acdonald conjecture and {H}odge theory on the
              loop {G}rassmannian'',
   Ann.\ of Math.\ {\bf 168}, 175 (2008).

\bibitem[GK25]{Gaiotto:2023dvs}
D.~Gaiotto and A.~Khan,
``Categorical pentagon relations and Koszul duality,''
Lett. Math. Phys. \textbf{115}, no.5, 100 (2025)
doi:10.1007/s11005-025-01932-1
[arXiv:2309.12103 [hep-th]].

\bibitem[GMW15]{Gaiotto:2015aoa}
D.~Gaiotto, G.~W.~Moore and E.~Witten,
``Algebra of the Infrared: String Field Theoretic Structures in Massive $\mathcal{N}=(2,2)$ Field Theory In Two Dimensions,''
[arXiv:1506.04087 [hep-th]].

\bibitem[Ko03]{Kontsevich:1997vb}
M.~Kontsevich,
``Deformation quantization of Poisson manifolds. 1.,''
Lett. Math. Phys. \textbf{66}, 157-216 (2003)
doi:10.1023/B:MATH.0000027508.00421.bf
[arXiv:q-alg/9709040 [math.QA]].

\bibitem[M10]{moorelectures}
G.~W.~Moore,
``PiTP Lectures on Wall-Crossing,''
Available at \url{https://www.physics.rutgers.edu/~gmoore/PiTP-LectureNotes.pdf}.

\bibitem[Niu22]{Niu:2021jet}
W.~Niu,
``Local operators of $4d \mathcal{N}=2$ gauge theories from the affine Grassmannian,''
Adv. Theor. Math. Phys. \textbf{26}, no.9, 3207-3247 (2022)
doi:10.4310/ATMP.2022.v26.n9.a10
[arXiv:2112.12164 [math.AG]].

\bibitem[Jeo19]{Jeong:2019pzg}
S.~Jeong,
``SCFT/VOA correspondence via $\Omega$-deformation,''
JHEP \textbf{10}, 171 (2019)
doi:10.1007/JHEP10(2019)171
[arXiv:1904.00927 [hep-th]].

\bibitem[Kac15]{kacreview}
V.~Kac,
``Introduction to vertex algebras, Poisson vertex algebras, and integrable Hamiltonian PDE,''
[arXiv:1512.00821 [math-ph]].

\bibitem[Kap06]{Kapustin:2006hi}
A.~Kapustin,
``Holomorphic reduction of N=2 gauge theories, Wilson-'t Hooft operators, and S-duality,''
[arXiv:hep-th/0612119 [hep-th]].

\bibitem[KZ25]{Khan:2025rah}
A.~Z.~Khan and K.~Zeng,
``Poisson Vertex Algebras and Three-Dimensional Gauge Theory,''
[arXiv:2502.13227 [hep-th]].

\bibitem[OY19]{Oh:2019bgz}
J.~Oh and J.~Yagi,
``Chiral algebras from $\Omega$-deformation,''
JHEP \textbf{08}, 143 (2019)
doi:10.1007/JHEP08(2019)143
[arXiv:1903.11123 [hep-th]].

\bibitem[OY20]{Oh:2019mcg}
J.~Oh and J.~Yagi,
``Poisson vertex algebras in supersymmetric field theories,''
Lett. Math. Phys. \textbf{110}, no.8, 2245-2275 (2020)
doi:10.1007/s11005-020-01290-0
[arXiv:1908.05791 [hep-th]].

\bibitem[SW94]{Seiberg:1994rs}
N.~Seiberg and E.~Witten,
``Electric - magnetic duality, monopole condensation, and confinement in N=2 supersymmetric Yang-Mills theory,''
Nucl. Phys. B \textbf{426}, 19-52 (1994)
[erratum: Nucl. Phys. B \textbf{430}, 485-486 (1994)]
doi:10.1016/0550-3213(94)90124-4
[arXiv:hep-th/9407087 [hep-th]].

\bibitem[Vaf98]{Vafa:1991uz}
C.~Vafa,
``Topological mirrors and quantum rings,''
AMS/IP Stud. Adv. Math. \textbf{9}, 97-120 (1998)
[arXiv:hep-th/9111017 [hep-th]].

\bibitem[WW24]{Wang:2024tjf}
M.~Wang and B.~R.~Williams,
``Factorization algebras from topological-holomorphic field theories,''
[arXiv:2407.08667 [math-ph]].

\bibitem[Wit07]{Witten:2005px}
E.~Witten,
``Two-dimensional models with (0,2) supersymmetry: Perturbative aspects,''
Adv. Theor. Math. Phys. \textbf{11}, no.1, 1-63 (2007)
doi:10.4310/ATMP.2007.v11.n1.a1
[arXiv:hep-th/0504078 [hep-th]].

\end{thebibliography}
\end{document}